\documentclass[journal,romanappendices]{IEEEtran}

\usepackage{algorithmicx}
\usepackage[ruled]{algorithm}
\usepackage{algpseudocode}
\usepackage{cite}
\usepackage[english]{babel}
\usepackage[latin1]{inputenc}
\usepackage{epsfig}
\usepackage{color}

\usepackage{times} 
\usepackage{subfigure}
\usepackage{url}
\usepackage{amsthm}
\usepackage{amsmath}
\usepackage{amssymb}
\usepackage{dsfont}
\usepackage[normalem]{ulem}

\usepackage{lipsum}
\usepackage{graphicx}

\graphicspath{{./Figuras/}}                   

\theoremstyle{definition} 
\newtheorem{corollary}{Corollary} 
\newtheorem{lemma}{Lemma} 

\newtheorem{remark}{Remark} 
\newtheorem{prop}{Properties} 

\newcommand{\prob}{\mathbb{P}}
\newcommand{\Ex}{\mathbb{E}}
\newcommand{\Var}{\mathbb{V}\mathrm{ar}}

\newcommand{\R}{\mathbb{R}}
\newcommand{\C}{\mathbb{C}}

\newcommand{\bigone}{\mathds{1}} 
\newcommand{\defeq}{\triangleq}
\newcommand{\Hip}{\mathcal{H}}
\newcommand{\N}{\mathcal{N}}
\newcommand{\CN}{\mathcal{CN}}

\newcommand{\tr}{\text{tr}}
\newcommand{\ve}[1]{\boldsymbol{#1}}
\newcommand{\mat}[1]{\mathbf{#1}}

\newif\ifshowfig 
\showfigtrue     

\begin{document}
\title{An Asymptotically Equivalent GLRT Test for Distributed Detection in Wireless Sensor Networks}

\author{{Juan Augusto Maya, Leonardo Rey Vega, Andrea M. Tonello~\IEEEmembership{Senior Member,~IEEE.}}
\thanks{J. A. Maya and A. M. Tonello are with the University of Klagenfurt, Klagenfurt, Austria. L. Rey Vega is with Universidad de Buenos Aires and CSC-CONICET, Buenos Aires, Argentina. The work of J. A. Maya has been supported by the Ubiquitous Sensing Lab, a joint laboratory between the University of Klagenfurt and Silicon Austria Labs.
Emails: \{juan.maya, andrea.tonello\}@aau.at, lrey@fi.uba.ar. 
} 
} 
\maketitle
\begin{abstract}
In this article, we tackle the problem of distributed detection of a radio source emitting a signal. We consider that geographically distributed sensor nodes obtain energy measurements and compute cooperatively a statistic to decide if the source is present or absent. We model the radio source as a stochastic signal and work with spatially statistically dependent measurements. We consider the Generalized Likelihood Ratio Test (GLRT) approach to deal with an unknown multidimensional parameter from the model. We analytically characterize the asymptotic distribution of the statistic when the amount of sensor measurements tends to infinity. Moreover, as the GLRT is not amenable for distributed settings because of the spatial statistical dependence of the measurements, we study a GLRT-like test where the statistical dependence is completely discarded for building this test. Nevertheless, its asymptotic performance is proved to be identical to the original GLRT, showing that the statistical dependence of the measurements has no impact on the detection performance in the asymptotic scenario. Furthermore, the GLRT-like algorithm has a low computational complexity and demands low communication resources, as compared to the GLRT.
\end{abstract}


\section{Introduction}
\subsection{Motivation and related work}
In recent years, wireless sensor networks (WSN) have received a lot of attention from the industrial and research community because of their remote sensing and control capabilities \cite{shaikh2016energy, rashid2016applications,li2020distributed}. More recently, they have become a crucial part of the emerging technology of Internet of Things (IoT) \cite{gupta2020collaborative,gubbi2013internet,al2015internet}. Among several topics related to WSNs, distributed detection is an active research area \cite{reisi2013distributed, tavana2016cooperative, chepuri2016sparse, aldalahmeh2019fusion,varshney2020distributed,leonard2018robust}.
In distributed detection problems, a set of nodes sense the environment in search for the presence of a source signal, which is typically linked with some physical process extended
over the geographical area where the network is deployed. Through collaboration among the nodes, and possibly with a fusion center (FC), the network is expected to decide with high confidence if the above-mentioned signal is present or not. A well-studied application example of this general problem is spectrum sensing \cite{Lunden_Koivunen_Poor_2015}.

Distributed detection architectures can be broadly classified in two classes. In the first class, all nodes transmit their local measurements (or a statistic of them) to a FC, where some processing tasks are done and the final decision about the underlying phenomenon is made \cite{blum1997distributed, viswanathan1997distributed, ChamberlandVeeravalli2003, varshney2012distributed,varshney2020distributed}. The communication between the nodes and the FC could be through possibly a multiple access channel (MAC), or a parallel access channel (PAC). In MAC channels, all nodes transmit synchronously and simultaneously a local statistic in order to build the final (global) statistic in the FC, to then make a decision \cite{Maya_2015,aldalahmeh2022distributed,banavar2012effectiveness}. On the other hand, in PAC channels the nodes communicate their local statistics through orthogonal channels with the FC \cite{LiDai_DetSignalMAC,SayeedTBMA,ciuonzo2021distributed}. Other authors have also considered clustered sensor networks, where the nodes are grouped into clusters, and the communication with the FC is through multiple hops \cite{sun2017multi,sah2020renewable,li2022distributed}. 
An alternative to the above architectures is to consider \emph{in-network processing} (IN-P) strategies without a FC. Here, the nodes build a local statistic, exchange information with their neighbors and, finally, execute some consensus or diffusion algorithm to achieve a common final decision \cite{bajovic2011distributed,xiao2004fast}. In this work, we do not restrict ourselves to any of these strategies, and we assume that the nodes cooperate using any of them. In addition, although distributed detection architectures typically consider one-bit or multi-bit quantization schemes at the nodes, we here assume that the nodes transmit their analog (unquantized) local statistic as in \cite{LiDai_DetSignalMAC,SayeedTBMA}, and leave the quantization problem for future works.

Distributed detection of a non-cooperative source emitting a radio signal has received a lot of attention in different scenarios.  
Some works considered fast-fading in the sensing channels (the channels that link the radio source with the nodes) \cite{maya2023selection,digham2003energy,choi2013soft}, while others have considered slow/block-fading channels \cite{maya2021effect, maya2021distributed, maya2021fully, ghasemi2005collaborative}. The first case is a representative of high-dynamic environments, as in vehicular networks \cite{peng2020enabling, balti2020mmwaves}, while the second one models slow-varying or stationary environments. 
On the other hand, the source detection problem can be modeled in different ways. Some authors modeled the radio source as a deterministic signal with an unknown \emph{scalar} parameter \cite{ciuonzo2021distributed, cheng2019multibit, ciuonzo2017distributed, ciuonzo2017generalized}. Under this model, and conditioned to whether the source is active (hypothesis $\Hip_1$) or not (hypothesis $\Hip_0$), the measurements at different nodes are statistically independent. In that setup, it is shown that locally most powerful tests or generalized score tests can be implemented. A generalization for the case with multiple unknown parameters was done in \cite{novikov2011locally} considering a max/min approach, where the probability density function (PDF) of the measurements belongs to the so-called locally asymptotic Gaussian families.

In each of the strategies (with or without FC), a crucial step for inexpensive WSNs is to build local statistics at each sensor site with minimum network resources (e.g., energy and bandwidth), and then implement a detector as a function of them. It is well-known that the optimal test in the Neyman-Pearson setting is the likelihood ratio test (LRT) \cite{Levy_Det}, which is defined through the quotient between the PDF of the measurements under the hypotheses $\Hip_1$ and $\Hip_0$. When the observations of different nodes are independent under both hypotheses, the PDF of the measurements is conveniently factorized and the local statistics are straightforwardly identified \cite{viswanathan1997distributed}. However, if the radio signal is modeled as a stochastic process, the node's observations are statistically dependent, given that each sensor receives a noisy version of the same source signal propagated through the corresponding wireless channel. In this case, no general factorizations of the involved PDFs are available, and obtaining the local statistics is challenging \cite{maya2022exponentially}. 

In general, distributed detection problems deal with sensors that acquire statistically independent observations \cite{sayed2014adaptation, sayed2013diffusion, ciuonzo2021distributed}. However, the scenario with spatially statistically dependent measurements can be extremely challenging \cite{willett2000good} and it is a partially explored area. In \cite{maya2021distributed, maya2021fully} a statistic based on the marginal PDFs of the measurements is analyzed for a general parametric model possibly considering statistically dependent observations with unrestricted parameters. In \cite{maya2021effect, maya2022exponentially} a source detection problem is discussed for sensors measuring the energy of a temporally-correlated Gaussian signal where the joint PDF of the measurements is not available when the source is active. In \cite{willett2000good} optimal quantization schemes are discussed for detecting a mean shift in a set of spatially correlated Gaussian observations. In \cite{zhang2020distributed} a copula-based distributed sequential detection scheme is designed to take the observations' spatial dependence into account. Different from previous works, we investigate the impact of the observations' spatial statistical dependence on the performance of a source detection problem for a particular physically-grounded model.

\subsection{Contributions and paper organization}

We assume that the nodes implement energy detectors, and deal with a model whose parameters are typically unknown, as for example, the energy of the radio source signal and the wireless sensing channel amplitudes. Then, it is shown that these parameters are restricted to be positive. This fact modifies the procedure for building the local statistics at the nodes, and also the theoretical characterization of detectors under analysis. This model is presented in Section II.

Since an optimum approach is not available, we follow the philosophy of the generalized likelihood ratio test (GLRT).
This approach is sometimes preferred because it has been proved to be optimal in terms of the error exponents for some particular distributions and settings \cite{Levy_Det}. 
The GLRT uses the joint PDF of the observations when the source signal is on and off
to build the test, so the statistical dependence of the observations is completely considered. However, as we will discuss in Sec. IV, its implementation is not amenable to WSNs with scarce resources such as energy and bandwidth. 

To sort this issue out, we study an algorithm that uses the product of the marginal PDFs as a substitute of the joint PDF for implementing a GLRT-like test. This strategy, called L-MP\footnote{The acronym L-MP refers to that the GLRT-like statistic estimates its parameter/s locally (L) and the joint PDF is replaced by the product of the marginal PDFs (MP) under each hypothesis}, allows us to immediately identify the local statistics at each sensor site for implementing a suitable  distributed detector. This statistic was proposed in \cite{maya2021distributed} for a general parametric model, and its performance was characterized for unrestricted estimators. Those results do not apply directly to the particular problem tackled here, given that the positivity constraints on the model's parameters poses a different and more challenging problem. Moreover, and differently from \cite{maya2021distributed}, we here are able to obtain the L-MP statistic in closed-form, which makes its computation more efficient at each sensor node.

Most importantly, the key point is that the L-MP strategy preserves the asymptotic performance of the GLRT. To prove this, in Section III, we first compute the asymptotic distribution of the GLRT under both hypotheses. When the unknown parameters have no constraints, the asymptotic distribution of the GLRT when the signal is off (on, resp.) is the well-known central (non-central) chi-square distribution. However, the positivity restriction on the estimated multidimensional parameters modifies the asymptotic distributions under both hypotheses, and it is carefully considered in the presented results.
Then, in Section IV, we also compute the asymptotic distributions of the L-MP detector under both hypotheses considering also the parameter constraints and show that they are exactly the same to those of the restricted GLRT.
In Section V, we conduct some Monte Carlo simulations and show that the L-MP detector has practically the same performance as the GLRT also when the data size at each sensor is relatively small. In addition, its asymptotic theoretical performance characterization presents only a small gap in that scenario. The main contributions of the paper can be summarized as follows:
\paragraph{Theoretical performance analysis} New theoretical results for characterizing the asymptotic performance of the GLRT and L-MP for the problem at hand are derived, for which the estimation of a multidimensional parameter with restrictions is needed. See Lemmas 1 and 2.

\paragraph{Restricted L-MP detector} A new computationally low-cost and low-communication resource demanding distributed detector is presented. The algorithm is implemented using only elementary functions, given that a closed-form expression is obtained for estimating the unknown parameters. In addition, and more importantly, it preserves the same asymptotic performance of the GLRT. See Section IV-A.

\paragraph{Impact of the statistical dependence of the observations} The theoretical results show that the sensor measurements' statistical dependence has no impact on the detector performance in the asymptotic scenario. This conclusion relies on the fact that the L-MP detector, based on the product of the marginal PDFs of the observations, completely discards the data dependence structure of the PDFs. However, it asymptotically achieves the same performance as the GLRT, which considers the joint PDFs instead. See Section IV-A, in particular, Remark 1.

We finally draw the conclusions of the work in Section VI, and relegate most of the technical details to the appendices.    

\begin{figure*}
	\centering
	\includegraphics[width=.9\linewidth]{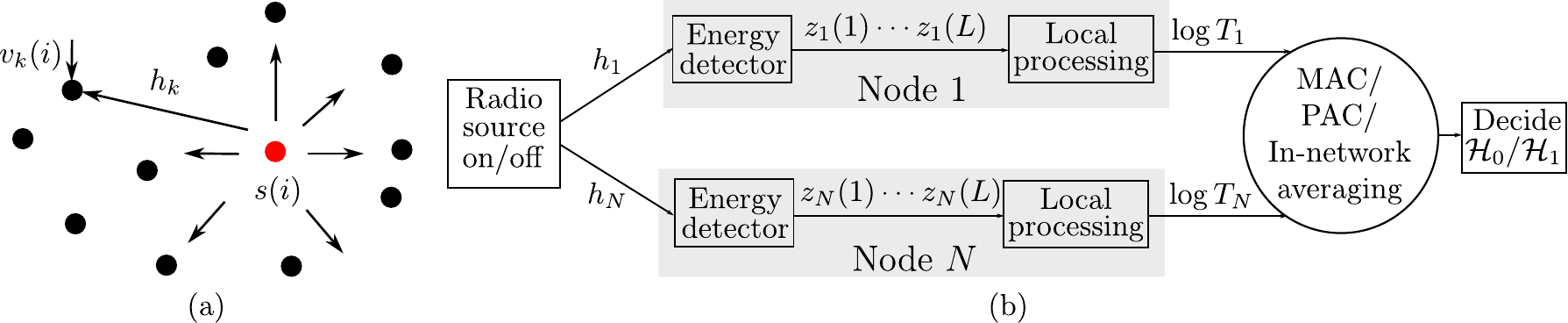}
	\caption{(a) Representation of the WSN for sensing the source's state. Nodes (black dots) distributed in space want to detect the presence or absence of a radio source (red dot). (b) Flowchart of the distributed detection system. Nodes observe energy measurements and generate local statistics which then are combined in a FC through a MAC/PAC channel or using some consensus algorithm to make the decision about the state of the source.}
	\label{fig:network}
\end{figure*}

\subsection{Notation}

We denote with $\mathbf{0}_N$ the $N$-dimensional null vector and with $\mathbf{I}_N$ the $N\times N$ identity matrix. 
Given a vector $\mathbf{a}$ we denote with $\text{diag}(\mathbf{a})$ a diagonal matrix with the diagonal entries given by the components of $\mathbf{a}$. 
With $\N(\ve \mu,\ve \Sigma)$ we denote a multivariate normal distribution with mean vector $\ve\mu$ and covariance matrix $\ve\Sigma$. With $\CN(0,\sigma^2)$ we denote the zero-mean complex circularly-symmetric normal distribution with variance $\sigma^2$. Given two $N$-dimensional vectors $\mathbf{a}$ and $\mathbf{b}$ we write $\mathbf{a}\succeq\mathbf{b}$ ($\mathbf{a}\succ\mathbf{b}$) if $a_i\geq b_i$ ($a_i>b_i$) for all $i\in[1,N]$.
$\{a\}_+\defeq \max(0,a)$, and $\{\ve a\}_+$ operates identically as $\{a\}_+$ over each component of $\ve a$. $\bigone(C)$ is the indicator function, i.e., $\bigone(C) = 1$ if the condition $C$ is true, and $0$ otherwise. We use the big-O notation for convergence in probability, that is, $f(x)=\mathcal{O}_p(g(x))$ for a strictly positive function $g(x)$ if $|f(x)|/g(x)\rightarrow K$ in probability, when $x \rightarrow\infty$ for a positive constant $K$.

\section{Model}
\label{sec:model}
We consider a WSN with $N$ nodes with sensing capabilities distributed in a bounded geographical area. We assume that at an unknown position there is a possibility of having a source emitting a signal $s(i)$. Through the processing of their observations, the network looks for the correct decision regarding the presence or absence of the source. See Fig. \ref{fig:network} for a schematic representation of the problem.
We assume that the nodes implement energy detectors, which is a common assumption for a WSN given that low-cost sensors are able to implement them. In addition, energy detectors can be implemented even in scenarios with a non-cooperative source, where the signal waveform is typically unknown. The energy detectors measure the energy of the signal during $T_0$ seconds. If $W$ is the sensed bandwidth, and according to \cite{landau1962prolate}, \cite[Ch. 8.4]{gallager1968information}, the energy detector measurements can be modeled as the energy of\footnote{Note that $M$ needs to be a positive integer. This is satisfied considering the largest integer smaller or equal to $WT_0$, i.e., $\lfloor WT_0\rfloor$.} $M\approx WT_0$ complex-baseband signal samples taken in a time window of duration $T_0$, as shown in (\ref{eq:SUenergy}). Let $y_n(i)$ denote the $i$-th signal sample, $i\in[1,M]$, taken by the $n$-th node. Under $\Hip_0$, the source is assumed to be silent and only thermal noise is present. Under $\Hip_1$, the source is assumed to be active and the received signal (before the energy detector) is the sum of thermal noise and the source signal:\footnote{We remark that the signal $y_n(i)$ is not available at each node. It is used only to model the energy measurements (\ref{eq:SUenergy}) that each receiver delivers every $T_0$ seconds.} 
\begin{equation}
\left\{\hspace{-1mm}
\begin{array}{ll}
\Hip_0: y_n(i)=v_n(i), \ i\in[1,M], n\in[1,N],\\
\Hip_1: y_n(i)=h_n s(i) + v_n(i).
\end{array}\right.
\label{eq:signal_rx}
\end{equation}
$h_{n}$ denotes the complex wireless channel gain from the source to the $n$-th node, $s(i)$ is assumed to be i.i.d. with distribution $s(i)\sim\CN(0,E_s)$, and $v_n(i)$ is the thermal noise sample at the $n$-th node, distributed as $v_n(i)\sim \CN(0,N_0)$, independent of $s(i)$. We assume that the source, the nodes and the environment are static during the observation time interval, thus, $h_{n}$ is constant during that time. That is, we consider a slow-fading scenario. The noise variance $N_0$ can be estimated in known silent time periods of the source, and it is assumed to be known at each node. The energy detector of the $n$-th node delivers the normalized energy level\footnote{We define the energy samples in this way in order to have zero-mean, unit-variance random variables under $\Hip_0$, as can be seen in App. \ref{app:gauss_mom}.}, that is:
\begin{equation}
z_n=\frac{1}{\sqrt{M}N_0}\sum_{i=1}^{M}(|y_n(i)|^2-N_0),\quad n \in[1,N]. \label{eq:SUenergy}
\end{equation}
Now we compute the joint PDF of the energy vector $\ve z \defeq [z_1,\dots,z_N]^T$. For many applications, the time-bandwidth product $M$ is typically $M\gg 1$, therefore the multidimensional central limit theorem (CLT) \cite{Billingsley_1995} can be used to approximate the joint distribution of $\ve z$ by a multivariate Gaussian distribution. Assuming that each node acquires $L$ independent and identically distributed (i.i.d.) energy measurements according to (\ref{eq:SUenergy}) during the observation time interval $LT_0$, i.e. $\ve z_{1:L} \defeq (\ve z_1\dots,\ve z_L)$, the hypothesis testing problem can be formulated as follows (see App. \ref{app:gauss_mom} for the details):
\begin{equation}
\left\{
\begin{array}{llll}
\Hip_0:& \ve z_{1:L} \overset{iid}{\sim} p(\ve z;\ve\theta_0) \\
\Hip_1:& \ve z_{1:L} \overset{iid}{\sim} p(\ve z;\ve\theta_1),
\end{array}\right.
\label{eq:htpeq}
\end{equation}
where the vector parameter $\ve \theta = [\theta_1,\dots,\theta_N]^T$ is a $N$-dimensional vector with positive or zero components as it is evident in App. \ref{app:gauss_mom}, i.e. $\ve \theta\succeq \ve 0_N$, $\ve \theta_0= \ve 0_N$, $\ve \theta_1=[E_s|h_1|^2/N_0,\dots,E_s|h_N|^2/N_0]^T$, and $p(\ve z;\ve\theta)= \N(\ve\mu(\ve \theta),\mathbf \Sigma(\ve \theta))$ with
\begin{equation}
\begin{array}{ll}
\ve \mu(\ve \theta) &= \sqrt{M}\ve \theta\\
\mathbf \Sigma(\ve \theta) & = \ve \theta \ve \theta^T + 2 \text{diag}(\ve \theta)  + \mathbf{I}_N.
\end{array}
\label{eq:meanCoveq}
\end{equation}
Under $\Hip_1$, the true vector parameter $\ve \theta_1$ is assumed to be unknown. This is because of the lack of knowledge of the true position of the source and its transmitted energy $E_s$, and because of the wireless channel amplitude $|h_n|$ is typically unknown. The wireless channel is influenced by several complex phenomena (e.g. path-loss, shadowing and fading) which are difficult to know in advance. Therefore, (\ref{eq:htpeq}) is a composite hypothesis testing problem, for which the vector parameter $\ve\theta_1\in\R^N_{>0}$ is unknown. Moreover, and to the best of the authors knowledge,  the optimal uniformly most powerful test (UMPT) is unknown or does not exist. This is a typical situation for composite hypothesis testing problems with multiple unknown parameters. Thus, we need a statistic that avoids the use of the unknown parameters or estimates them in some way. In this paper, we follow the approach of the GLRT, which estimates the unknown vector parameter using the MLE. Observe that techniques as locally most powerful test \cite[Sec. 6.7]{Kay_SSP} or locally directionally maximin test \cite{novikov2011locally} cannot be applied for different reasons. The former one is only valid for PDFs with a scalar parameter, and the latter one is only valid for the so-called locally asymptotically normal families. Unfortunately, the PDF (\ref{eq:htpeq}) under $\Hip_1$ does not belong to that class.       

\section{The restricted GLRT}
The GLRT is a well-known technique frequently used for composite hypothesis testing problems, like the one presented in the previous section. Its success relies on the asymptotic properties of the MLE, which, under mild conditions, is an efficient estimator when the number of observations tends to infinity \cite{Kay_SSP_ET}. In addition, the GLRT is proved to be optimal in terms of the error exponents for several distributions, including the Gaussian distribution and the exponential families of probability distributions \cite[Sec. 5.6.2]{Levy_Det}, \cite[Th. 3.2.1, Th. 3.2.2]{kallenberg1978asymptotic}. Moreover, under some mild assumptions, and when there are no constraints on the unknown parameters, the exact asymptotic characterization of the GLRT is easily obtained \cite{Kay_SSP}. On the other hand, in our problem we do have constraints: the entries of $\ve \theta_1$ are nonnegative. This fact impacts on the asymptotic distributions of the GLRT under both hypotheses, but only partial results can be found in the current literature. We next present a complete characterization of the asymptotic distribution of the GLRT, which allows us to analytically compute its error probabilities when the amount of observations $L$ tends to infinity.

Let $f(\ve z_{1:L},\ve \theta)\defeq \prod_{l=1}^{L} p(\ve z_l,\ve\theta)$ be the likelihood function with parameter $\ve\theta$. The GLRT statistic is defined as \cite{chernoff1954distribution}:
\begin{equation}
T_\text{G}(\ve z_{1:L})\defeq \frac{\max_{\ve \theta\succeq\ve 0_N} f(\ve z_{1:L},\ve \theta)}{f(\ve z_{1:L},\ve 0_N)}. \label{eq:glrt}
\end{equation}
Under the model (\ref{eq:htpeq}), it evaluates as follows:
\begin{align}
\tfrac{2}{L}\log T_\text{G}(\ve z_{1:L})&= -\log\det\mat{\Sigma}(\hat{\ve\theta})+\tfrac{1}{L}\sum_{l=1}^L \|\ve z_l\|^2 \nonumber\\
&- (\ve z_l-\ve\mu(\hat{\ve\theta}))^T \mat{\Sigma}^{-1}(\hat{\ve\theta}) (\ve z_l-\ve\mu(\hat{\ve\theta})),
\label{eq:glrt_gauss}
\end{align}
where $\hat{\ve\theta}$ is the positive constrained \emph{global} MLE\footnote{We call this estimator \emph{global} MLE to differentiate it from the local one $\hat {\ve \theta}_{\text{L-MLE}}$ to be defined in the following section.} defined as
\begin{align}
	\hat{\ve\theta} &\defeq \arg\max_{\ve \theta\succeq\ve 0_N} f(\ve z_{1:L};\ve\theta)\nonumber\\
	&= \arg\min_{\ve \theta\succeq\ve 0_N}\log\det\mat{\Sigma}(\ve\theta)\nonumber\\ &+\tfrac{1}{L}\sum_{l=1}^L (\ve z_l-\ve\mu({\ve\theta}))^T \mat{\Sigma}^{-1}({\ve\theta}) (\ve z_l-\ve\mu({\ve\theta})).
	\label{eq:GMLE}
\end{align}

The optimization set of $\ve\theta$ in (\ref{eq:glrt}) and (\ref{eq:GMLE}) is the cone $\ve\Theta = \{\ve\theta\in\R^N:\ve \theta\succeq \ve 0_N\}$. The true value of $\ve\theta$ under $\Hip_0$, $\ve\theta_0=\ve 0_N$, is a corner point of $\ve\Theta$. As the true parameter under $\Hip_0$ is not an interior point of $\ve\Theta$, the usual asymptotic analysis (see for instance \cite{Kay_SSP}) for the GLRT under $\Hip_0$ is not valid. 
On the other hand, the true value of the parameter $\ve\theta$ under $\Hip_1$, $\ve\theta_1\succ\ve 0_N$, is an interior point of $\ve\Theta$. However, the optimization set also modifies the asymptotic distribution of the GLRT under the so-called \emph{weak} signal assumption, i.e., when $\ve\theta_1$ scales as $\tfrac{1}{\sqrt{L}}$ when $L\rightarrow\infty$. This condition is necessary in order to have a non-degenerated
hypothesis testing problem\footnote{A degenerated hypothesis testing problem is a problem for which $P_\text{md} = 0$ and $P_\text{fa} \leq \epsilon$, for any $\epsilon > 0$ simultaneously, when $L\rightarrow\infty$, using a simple test statistic as for example the sample mean.}.

For a given statistic $T$ and a given predefined threshold $t$, the false alarm probability and the miss-detection probability are defined, respectively, as $P_\text{fa}=\prob(T>t;\Hip_0)$ and $P_\text{md}=\prob(T\leq t; \Hip_1)$. Then, we can express $P_\text{fa}= 1-F_T^0(t)$ and $P_\text{md}=F_T^1(t)$ in terms of the cumulative distribution function (CDF) under $\Hip_0$, $F_T^0(t)$, and, under $\Hip_1$, $F_T^1(t)$.
We summarize the asymptotic performance characterization of the GLRT in the following lemma. We obtain a closed form expression for it under $\Hip_0$. Nevertheless, this is not possible under $\Hip_1$. In the latter case, we get the characteristic function of the GLRT under $\Hip_1$ instead. Then, the corresponding CDF can be computed using a numerical method for evaluating its inverse Fourier transform (e.g., see \cite{abate2006unified}).
We first enumerate the following properties.
\begin{prop}
	\label{lemma:1}
The PDF $p(\ve z;\ve\theta)$ in (\ref{eq:htpeq}) satisfies 
\begin{enumerate}
	\item[P1)] For almost all $\ve z$, the derivatives $\frac{\partial\log p}{\partial\theta_n}$, $\frac{\partial^2\log p}{\partial\theta_n \partial\theta_k}$ and $\frac{\partial^3\log p}{\partial\theta_n \partial\theta_k \partial\theta_m }$, $n,k,m = 1,\dots,N$, exist for every $\ve \theta$ in the closure of a neighborhood $\mathcal{B}$ of $\ve \theta =\ve 0$.
	\item[P2)] If $\ve\theta\in\mathcal{B}$, 
$\Ex\left|\frac{\partial^3\log p}{\partial\theta_n \partial\theta_k \partial\theta_m}\right|<A$, where $A$ is independent of $\ve \theta$. 
	\item[P3)] If $\ve\theta\in\mathcal{B}$, the Fisher information matrix $\ve i(\ve 0)$ with $i,j$-element $(\ve i(\ve 0))_{i,j} = \Ex\big(\frac{\partial\log p(\ve z,\ve 0)}{\partial \theta_i}\frac{\partial\log p(\ve z,\ve 0)}{\partial \theta_j} \big)$ is finite and positive definite.
\end{enumerate}
\end{prop}
\begin{proof}
	See App. \ref{app:prop_pdf}.
\end{proof}
\begin{lemma}
\label{lemma:2}
Considering that $p(\ve z,\ve\theta)$ satisfies the properties enumerated in Properties 1, and assuming that there exists a positive constant $c$ such that $\|\ve\theta_1\|\leq c/\sqrt{L}$ (weak signal assumption under $\Hip_1$), the asymptotic distribution of the GLRT (\ref{eq:glrt}), when $L$ is sufficiently large, is $2\log T_\text{G}(\ve z_{1:L})\overset{a}{\sim}$
\begin{equation}
\left\{\hspace{-1mm}
\begin{array}{l}
\sum_{n=1}^N \{u_{0,n}\}_+^2\ :\Hip_0,  \ u_{0,n}\sim \N(0,1), n\in[1,N], \\
\sum_{n=1}^N \{u_{1,n}\}_+^2\ :\Hip_1,  \ u_{1,n}\sim \N(\psi_n ,1),	
\end{array}\right.
\label{eq:chi2}			
\end{equation}
where $\overset{a}{\sim}$ means ``asymptotically distributed as", $\{u_{0,n}\}_{n=1}^N$ and $\{u_{1,n}\}_{n=1}^N$ are independent random variables, and $\psi_n \defeq \sqrt{L(M+2)} E_s|h_n|^2/N_0$, $n\in[1,N]$. The closed-form expression of the CDF of (\ref{eq:chi2}) under $\Hip_0$, $F_{T_\text{G}}^0(t)=\prob(2\log T_\text{G}(\ve z_{1:L})\leq t;\Hip_0)$, is
\begin{equation}
F_{T_\text{G}}^0(t)=
\left(\frac{1}{2^N} + \sum_{n=1}^{N}{{N}\choose{n}}\frac{1}{2^N} F_n(t)\right) \bigone(t\geq 0).
\label{eq:CDF_H0}
\end{equation}
$F_n(t)$ is the central chi-square CDF with $n$ degrees of freedom.
In addition, the characteristic function of (\ref{eq:chi2}) under $\Hip_1$ is 
\begin{align}
	\Psi^1_{T_\text{G}}(\omega) = \prod_{n=1}^N \Big\{ \Phi(-\psi_n) + \tfrac{1}{2} (1-2\jmath \omega)^{-\frac{1}{2}}  
	 e^\frac{\jmath \omega \psi_n^2}{1-2\jmath \omega} \nonumber\\
	 \times\left(1-\text{erf} \left(-\psi_n (2(1-2\jmath \omega))^{-\frac{1}{2}}\right) \right)\Big\}.
	 \label{eq:chi2trunc}
\end{align}
$\Phi$ is the CDF of a standard Gaussian random variable, $\jmath$ is the imaginary unit, and $\text{erf}:\C\rightarrow\C$ is the error function defined as $\text{erf}(z) \defeq \frac{2}{\sqrt{\pi}}\int_0^z e^{-u^2/2} du$.
\end{lemma}
\begin{proof} 
	See App. \ref{app:ad_glrt}.
\end{proof}
It is clear from (\ref{eq:chi2}) that the asymptotic distribution of the GLRT under $\Hip_0$ ($\Hip_1$, resp.) is the distribution of the sum of $N$ independent truncated central (non-central) chi-square random variables with one degree of freedom. The difference between these results and the classical results for the unrestricted GLRT is the truncation of the Gaussian random variables in (\ref{eq:chi2}) performed by the operator $\{\cdot\}_+$.

\begin{remark}
The asymptotic characterization in Lemma 1 is important because it provides theoretical guaranties, it allows us to understand the impact of each involved parameter in the GLRT performance, but also has a practical implication: the distribution $F_{T_\text{G}}^0(t)$ can be used to set the GLRT threshold for obtaining $P_\text{fa}\leq \xi$, where $\xi$ is the desired size of the test. We will use this in the numerical experiments in Section V.
\end{remark}
\section{A GLRT-equivalent distributed statistic}
\subsection{Algorithm design and performance characterization}
As we will discuss in Sections IV-B and V, the implementation of the GLRT in distributed settings demands relatively high communication resources (energy and bandwidth) and it has an elevated computational complexity, especially for WSNs with limited communications resources and computational capabilities. We look for a more efficient approach in terms of network resources to build a test suitable for distributed scenarios based on the following observation: if each node considers only its own measurements, thanks to the special structure of (\ref{eq:meanCoveq}), it can estimate its corresponding component of $\ve\theta$, and then build cooperatively a statistic with the others nodes. Specifically, we study the L-MP statistic based on the product of the marginal PDFs instead of the joint PDF in (\ref{eq:glrt}), where the parameter estimation at each node is done using only its local measurements. This statistic, noted hereafter as $T_\text{L-MP}$, is defined as follows for the problem at hand: 
\begin{equation}
T_\text{L-MP} (\ve z_{1:L}) \defeq \prod_{k=1}^{N}\frac{\max_{\theta_k\geq 0} f_k(z_k(1),\dots,z_k(L);\theta_k)}{f_k(z_k(1),\dots,z_k(L);0)}, \label{eq:LMP}
\end{equation}
where $f_k(z_k(1),\dots,z_k(L);\theta_k) = \prod_{l=1}^L p_k(z_k(l),\theta_k)$ and \begin{equation}
p_k(z_k(l),\theta_k)=\N(\sqrt{M} \theta_k,(\theta_k+1)^2)
\label{eq:marginal}
\end{equation}
is the marginal PDF of $p(\ve z_l,\ve\theta)$ for the component $z_k(l)$. We also define the local MLE of $\theta_k$, $k\in [1,N]$ as 
\begin{equation}
	\hat{\theta}_{\text{L-MLE},k} \defeq \arg\max_{\theta_k\geq 0} f_k(z_k(1),\dots,z_k(L);\theta_k).
	\label{eq:L-MLE}
\end{equation}
It is shown in App. \ref{app:MLEloc} that the local MLE has the following closed-form expression:
\begin{multline}
	\!\!\!\!\!\hat{\theta}_{\text{L-MLE},k}\!=\!\frac{1}{2}\left\{\!\sqrt{ (M\!+\!2\!+\!\sqrt{M} m^1_k)^2\! +\! 4(m^2_k\!+\!\sqrt{M} m^1_k-1)}\right.\\
	\left. - (M+2+\sqrt{M} m^1_k)\right\}_+ \ k\in[1,N],
	\label{eq:mlelocal}
\end{multline}
where $m^1_k \defeq \textstyle\frac{1}{L}\sum_{l=1}^{L} z_{k}(l)$, and $m^2_k\defeq\textstyle\frac{1}{L}\sum_{l=1}^{L} z_{k}^2(l)$. Then, (\ref{eq:LMP}) can be written as follows
\begin{align}
	\log T_\text{L-MP} (\ve z_{1:L}) &= \sum_{k=1}^N \log T_k,  \label{eq:LMP2}\\ 
	\log T_k &=
	\sum_{l=1}^L  -\log (1+\hat{\theta}_{\text{L-MLE},k}) + \frac{1}{2} z_k^2(l)\nonumber\\ 
	 &-\frac{1}{2} \left(\frac{z_k(l)-\sqrt{M}\hat{\theta}_{\text{L-MLE},k}} {1+\hat{\theta}_{\text{L-MLE},k}}\right)^2.
	\label{eq:local}
\end{align}
We next prove that the L-MP detector (\ref{eq:LMP2}) preserves the asymptotic performance of the GLRT.   
\begin{lemma}
\label{lemma:L-MP}
 The asymptotic distribution of $T_\text{G}$ and $T_\text{L-MP}$ are exactly the same under both hypotheses when $L$ grows unbounded.
\end{lemma}
\begin{proof}
See App. \ref{app:L-MP}
\end{proof}

\begin{remark}
The previous result can be explained as follows. As shown in (\ref{eq:glrt2}) and (\ref{eq:glrt3}) in App. \ref{app:ad_glrt}, the asymptotic distribution of the GLRT under both hypotheses $\Hip_0$ or $\Hip_1$ (using the weak signal assumption) depends on the Fisher information matrix under $\Hip_0$, $\ve i(\ve 0_N)$. Given that $\ve i(\ve 0_N)$ is a diagonal matrix, and its diagonal elements depends exclusively on the marginal PDFs of the observations, the statistical properties of the GLRT test in the asymptotic scenario is preserved by the L-MP test, which is based on the marginal PDFs. Therefore, both tests are asymptotically equivalent, and the statistical dependence of the measurements for the finite data regime (c.f. (\ref{eq:meanCoveq})) has no influence in the asymptotic scenario, when $L\rightarrow\infty$.   
\end{remark}

\subsection{Communication resources for the GLRT and L-MP}
In Table \ref{tab:resources}, we compare the communication resources (energy and bandwidth) needed for implementing both the GLRT and the L-MP tests. The energy and the bandwidth resources are estimated as the number of transmissions and channel uses, respectively. We consider three typical communication scenarios: parallel access channel, multiple access channel and in-network processing. In all cases, we assume that the communication is through error-free channels. 

As seen in (\ref{eq:glrt_gauss}), the implementation of the GLRT requires, on the one hand, to compute $L$ quadratic forms of the vector $\ve z_l$, whose elements are the measurements of different nodes for the time slot $l\in[1,L]$. On the other hand, it also requires the computation of the global MLE (\ref{eq:GMLE}). This is a resource-demanding task that strongly depends on the network size $N$ and the number of samples at each node $L$, given that no closed-form expression is available for it and an iterative algorithm is required to solve the optimization problem with constraints. That is, there is no processing strategy for building cooperatively (with or without FC) and efficiently the GLRT. The naive approach of communicating all sensor measurements to the FC using a PAC channel requires $NL$ transmissions and channel uses. In the case of IN-P (without a FC), a flooding algorithm \cite{kshemkalyani2011distributed} can be used to communicate all measurements to the remaining nodes. In this case, the $N$ nodes transmit $N_f L$ messages, where $N_f$ depends on the network topology. $N_f$ is proportional to the number of edges of the graph that models the communication network, and it is typically $N_f>N$ \cite{kshemkalyani2011distributed}. A distributed algorithm designed for using minimal communication resources should employ substantially less transmissions than both previous strategies. In the case of the MAC channel, there is no straightforward strategy for implementing the GLRT and it is not considered in this case. 

The L-MP test requires the implementation of the statistic (\ref{eq:LMP2}). Its efficiency in terms of communication resources relies on the fact that each node is able to compute the \emph{local statistic} (\ref{eq:local}) without cooperating with other nodes for this task, given that, by definition in (\ref{eq:LMP}), it only depends on the locally acquired observations. 
Therefore, no communication resources are spent for this task. Then, only a cooperation step between the nodes (or with the FC, if present) is needed for computing (\ref{eq:LMP2}). In the case of a MAC channel (when there is a FC), and after some calibration step of the channels \cite{MayaBiasCorr}, the sensor node $k$ transmits $\log T_k$ simultaneously with the remaining sensor nodes to the FC, which receives (\ref{eq:LMP2}), if the communication noise is negligible. Thus, $N$ transmissions are needed but only one channel use is required, given that all nodes use the channel simultaneously. If the PAC channel is used instead, the local statistics $\{\log T_k\}$ are communicated through orthogonal channels, and then (\ref{eq:LMP2}) is implemented. This option requires $N$ transmissions and $N$ channel uses. Finally, in a fully-distributed scenario with in-network processing, (\ref{eq:LMP2}) can be computed with some consensus algorithm, as for example \cite{xiao2004fast}, where the nodes transmit messages only to their neighbors. In this case, $\beta N$ transmissions and channel uses are required, where $\beta$ is a constant that depends on the network topology and the consensus algorithm. Typically, $\beta<N$ \cite{xiao2004fast}. 

It is clear that the L-MP test is more efficient than the GLRT test when using the PAC and IN-P communication strategies, and it also could be implemented efficiently with a MAC channel. In the next section, we also show the computational benefits of the L-MP strategy.  

\begin{table}
	\centering
	\caption{Network communication resources for implementing the GLRT and L-MP tests.}
	\label{tab:resources} 	\begin{tabular}{|c|c|c|c|c|c|c|}
		\hline
		&  \multicolumn{3}{|c|}{GLRT} &  \multicolumn{3}{|c|}{L-MP}   \\
		\hline
		Netw. Arch.	& MAC &PAC 	& IN-P  		& MAC & PAC & IN-P \\
		\hline
		Energy		& $-$ & $NL$  & $N_f NL$ 	& $N$ & $N$ & $\beta N$ \\
		\hline
		Bandwidth	& $-$ & $NL$ 	& $N_f NL$ 	& $1$ & $N$ & $\beta N$ \\
		\hline
	\end{tabular}
\end{table}

\section{Results}
\begin{figure}[tb]
	\centering
	\includegraphics[width=.8\linewidth]{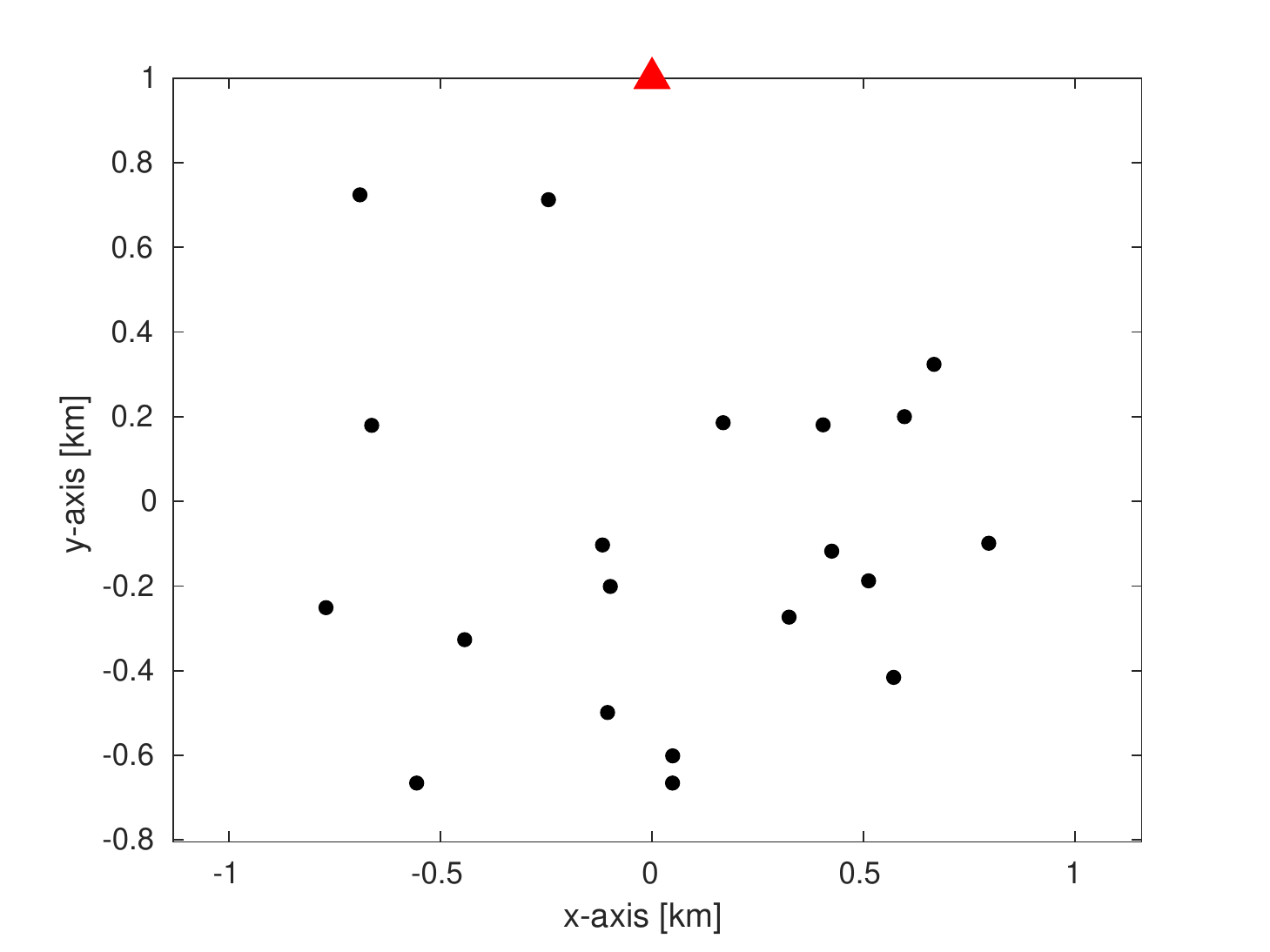}
	\caption{One realization of the network topology with $N=20$ nodes (black dots) and the radio source (red triangle) for the considered scenario. 
	}
	\label{fig:scenario}
\end{figure}

Next we evaluate the performance of the L-MP statistic and compare it against the GLRT and other detectors for finite-length data. The source is assumed to be a communication signal, as can be found in applications of spectrum sensing for IoT devices \cite{ansere2019reliable, dao2020energy, Lunden_Koivunen_Poor_2015}. In this section, we will consider two possible distributions for the  samples $s(i)$: i) the circular complex Gaussian distribution, as it was assumed for deriving the model in Sec. II; and ii) a $16$-QAM\footnote{We have also simulated other constellation orders obtaining similar results. We decided not to include them to keep the figures clear.} constellation with uniformly distributed symbols.
The numerical performance of each algorithm is computed using $10^5$ Monte Carlo runs. The data is generated following the model in (\ref{eq:SUenergy}), i.e., without considering that the distribution of the energy measurements is Gaussian. The channel variance $\sigma^2_{n}$, which is equal to the quotient between the received power (without considering the noise) and the transmitted power, is modeled using the path-loss/log-normal shadowing model \cite{goldsmith2005wireless},
\begin{equation}
	\sigma^2_{n}({\rm dB}) = K - 10\alpha \log_{10}(d_n/d_0) - \eta_n,\label{eq:var_h}	
\end{equation}
where $d_n$ is the distance between the source position and the $n$-th node position, $K$ (in dB) is the path-loss attenuation at a certain distance $d_0$, $\alpha$ is the path-loss exponent, and $\eta_{n}$ is a zero-mean Gaussian random variable with variance $\sigma^2_\eta$ which models the shadowing effect. In order to compute $d_n$, we assume that the node positions are uniformly randomly distributed in an square area of 1600 m$\times$1600 m, centered at $(0,0)$ m, and the position of the radio source is $(0,1000)$ m. In Fig. \ref{fig:scenario}, we show a realization of the sensor network topology. 

The channel gains are determined as follows. For each Monte Carlo run, $\sigma^2_n$ is sampled following (\ref{eq:var_h}). Then, as it is assumed that the source and the sensors are static, the channel gains $\{h_{n}\}_{n=1}^N$ (which are the same for all time windows $l\in[1,L]$) are i.i.d. sampled from the PDF $\CN(0,\sigma^2_n)$. That is, we assume that the channel amplitude is Rayleigh distributed, representative of a non-line-of-sight slow  fading channel. Therefore, with this procedure, and after considering all Monte Carlo runs, we are able to obtain the average performance of the detectors with respect to the channels' distribution (induced by the source-nodes distance distribution, and the path-loss/shadowing model).

\begin{table}[bt!]
	\caption{Selected parameters for the simulation setup.}
	\centering
	\resizebox{\columnwidth}{!}{\begin{tabular}{ccccccccccccc}
			$K$(dB) &  $\alpha$ & $d_0$(m) & $\sigma_\eta$ & $N_0$(dBm/Hz) & $N$ & $W$(MHz) \\
			\hline
			-37 &  4 & 10 & 3 & -174 & 20  & 5 \\ 
	\end{tabular}}
	\label{tab:param}
\end{table}
The chosen parameters for the simulation setup are shown in Table \ref{tab:param}. The selected propagation model parameters ($K$, $\alpha$, $d_0$ and $\sigma_\eta$) are typical for outdoors scenarios \cite[Ch. 2]{goldsmith2005wireless}.
The signal-to-noise ratio is defined by $\text{SNR}\defeq \frac{E_s\bar{\sigma}^2}{N_0}$, where $\bar{\sigma}^2$ is the average variance of the channels (\ref{eq:var_h}) across the nodes. Notice that the source energy $E_s$ is varied to be consistent with the corresponding SNR. 

We compare the performance of the L-MP algorithm against the ones shown in Table III (see the definitions of the acronyms in the third column). LR is the likelihood ratio test, which is the optimal test when the vector parameter $\ve\theta_1$ is perfectly known, which is not the case in most of the applications of interest. It is included only as a reference for comparison. Note that some performance loss of the GLRT-based algorithms is expected with respect to this \emph{genie-aided} test statistic due to the estimation errors of the unknown parameters.

The test statistic MD is the mean detector, i.e., the average of all network measurements, and is equivalent to the equal gain combining (EGC), typically used in low SNR regimes \cite{niu2006fusion}. The test statistic SD is the square detector. 
The SMC detector \cite{digham2003energy} selects the highest average measured energy among all the sensors. We also include two eigenvalue-based detectors (SSE and ME), which are typically used in similar scenarios. These detectors naturally consider the statistical dependence of the observations under $\Hip_1$ at different sensor nodes, introduced by the random signal source, given that they are based on the eigenvalues $\{\lambda_n\}_{n=1}^N$ of the sample covariance matrix of the observations $\{\ve{z}_l\}_{l=1}^L$. Nevertheless, judging by the theoretical analysis in the previous section, and the following numerical results, the statistical dependence appears to have a negligible impact in the detectors' performance in this scenario. 
Finally, we also consider the Rao test statistic (RD)\footnote{The statistic $1/L\tfrac{\partial \log p(\ve z_{1:L}; \ve 0)^T}{\partial\ve \theta} \ve i(\ve 0)^{-1} \tfrac{\partial\log p(\ve z_{1:L}; \ve 0)}{\partial\ve \theta}$ is the general expression for the Rao detector \cite{Kay_SSP}, and evaluates as it is shown in the table for the model at hand.}. 

In the last column of Table \ref{tab:stats}, we also include an estimation of the computational complexity of each detector by computing the average CPU time\footnote{One core of an Intel Xeon CPU ES-2690 v2 3.0 GHz server is used for running the simulations.} needed to implement the corresponding algorithm for each Monte Carlo run for SNR$=-10$ dB. Observe that the estimated complexity corresponds to the total complexity of the algorithms, which includes the local complexity at each sensor node. The computational cost of the L-MP test is four orders of magnitude smaller that the one of the GLRT. It is also less than the one of the eigenvalue-based algorithms (SSE and ME) and SMC, and one order of magnitude more than the remaining algorithms (MD, SD, and RD), whose performance is degraded with respect to the L-MP test, as we will see next.
The high computational cost of the GLRT is due to the numerical procedure needed for computing the positive constrained MLE of the $N$-dimensional parameter $\ve\theta_1$. We used here a trust-region algorithm for which the number of iterations needed depends on the
scenario (fundamentally on the SNR) and the stopping criteria parameters (e.g. the gradient tolerance and step tolerance). We used the default settings \cite{branch1999subspace}.

\begin{table}[t]
\caption{Test statistics to be compared. $\dagger$ is a genie-aided detector.}
\resizebox{\columnwidth}{!}{\begin{tabular}{|c|l|l|c|}
\hline
Label	&  Test statistic/Ref. &  Detector name/Observation & Avg. CPU \\ &&&time (us)\\
\hline
LR$\dagger$	& Eq. (\ref{eq:glrt_gauss}), with $\hat{\ve\theta}$ replaced by& Likelihood-ratio test detector. & $-$\\ 
& the true parameter $\ve\theta_1$.&& \\
\hline
GLRT	& Eq. (\ref{eq:glrt_gauss}) & GLRT detector. & 53921\\
\hline
L-MP	& Eq. (\ref{eq:LMP2})   &  Local marginal product detector. & 6.5\\
\hline
MD	& $1/(NL) \sum_{n,l} z_n(l)$&  Mean detector.& 0.4\\
\hline
SD	& $1/(NL) \sum_{n,l} z_n^2(l)$ &  Square detector.& 0.5\\
\hline
SMC & $\max_{n\in[1,N]} m^1_{n}$,\cite{digham2003energy}	& Selection maximum combining & 11.5\\ &&detector.&\\
\hline
SSE & $\sum_{n=1}^N -\log\lambda^+_n + \lambda^+_n$,\cite{zhang_multi-antenna_2010}	& Subspace eigenvalue detector. & 91.5 \\
&&$\lambda^+_n \defeq \{\lambda_n-1\}_+$ & \\
\hline
ME & $\max_{n\in[1,N]}\lambda_\text{n}$, \cite{taherpour_multiple_2010}	& Maximum eigenvalue detector.& 81\\
\hline
RD & $1/(NL)\sum_{k,l} (z_k(l)^2$ & Rao detector.& 0.5\\
& $+\sqrt{M}z_k(l)-1)^2$ \cite{Kay_SSP}&&\\
\hline
\end{tabular}}
\label{tab:stats}
\end{table}

\begin{figure}
	\centering
	\subfigure[$M=50$, $L=10$.] {\includegraphics[width=.85\linewidth]{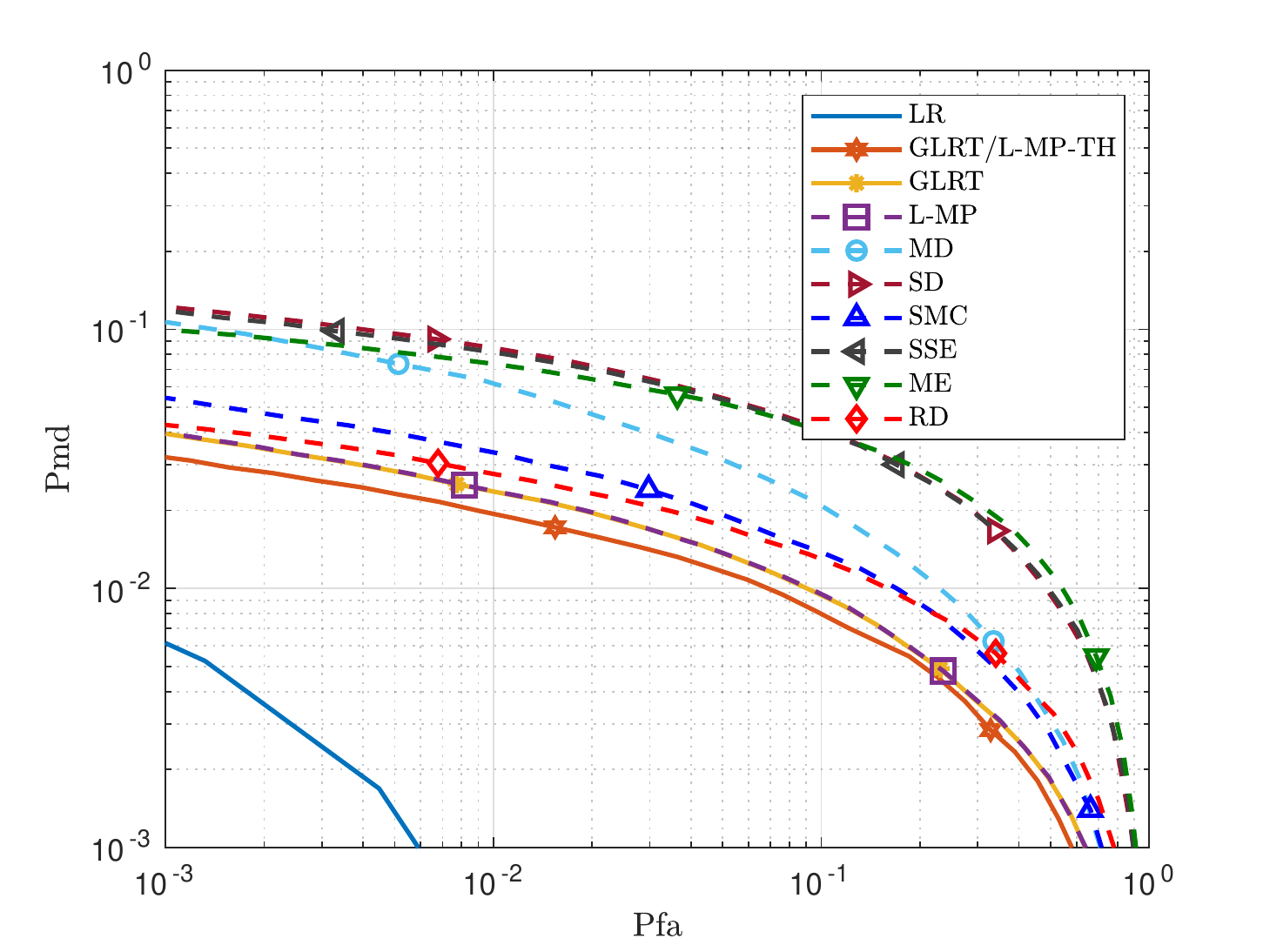}}
	\subfigure[$M=20$, $L=10$.] {\includegraphics[width=.85\linewidth]{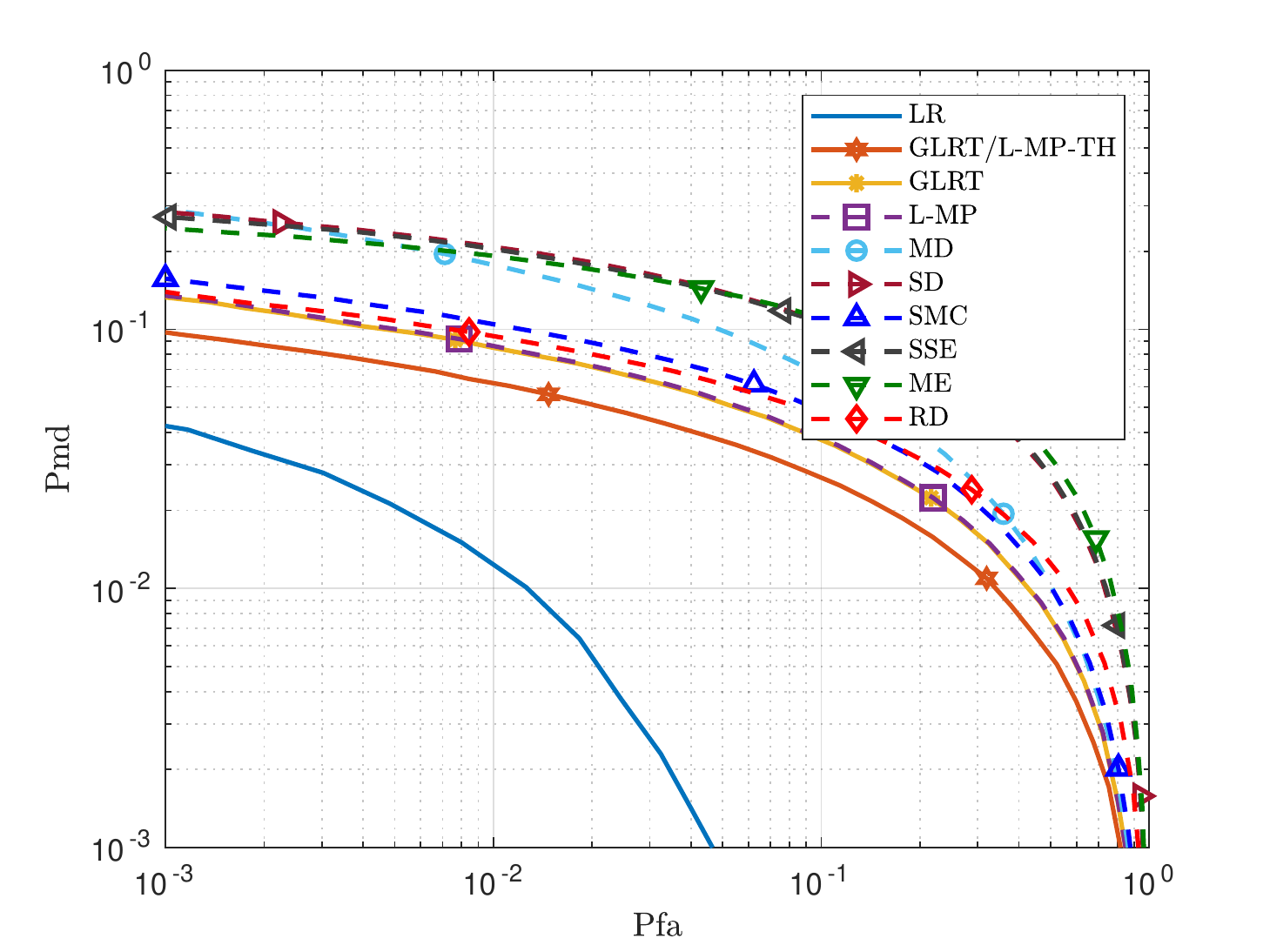}}
	\caption{Complementary receiver operating characteristics when the source is circularly-symmetric complex Gaussian distributed with the indicated $M$ and $L$, and $\text{SNR}=-11$ dB.}
	\label{fig:crocs}
\end{figure}

\begin{figure}
	\subfigure[$M=50$, $L=10$.] {\centering\includegraphics[width=.85\linewidth]{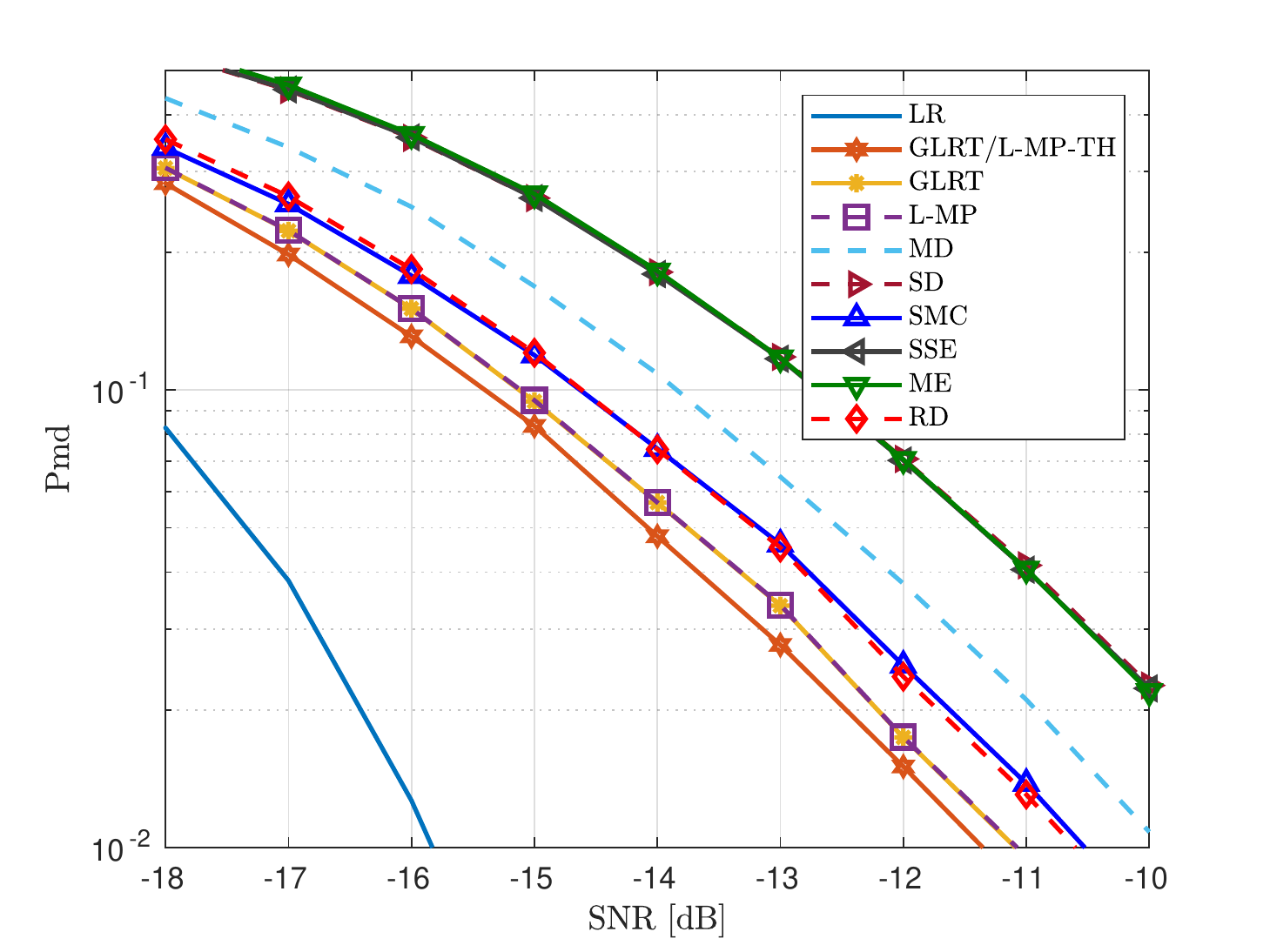}}
	\subfigure[$M=20$, $L=10$.] {\includegraphics[width=.85\linewidth]{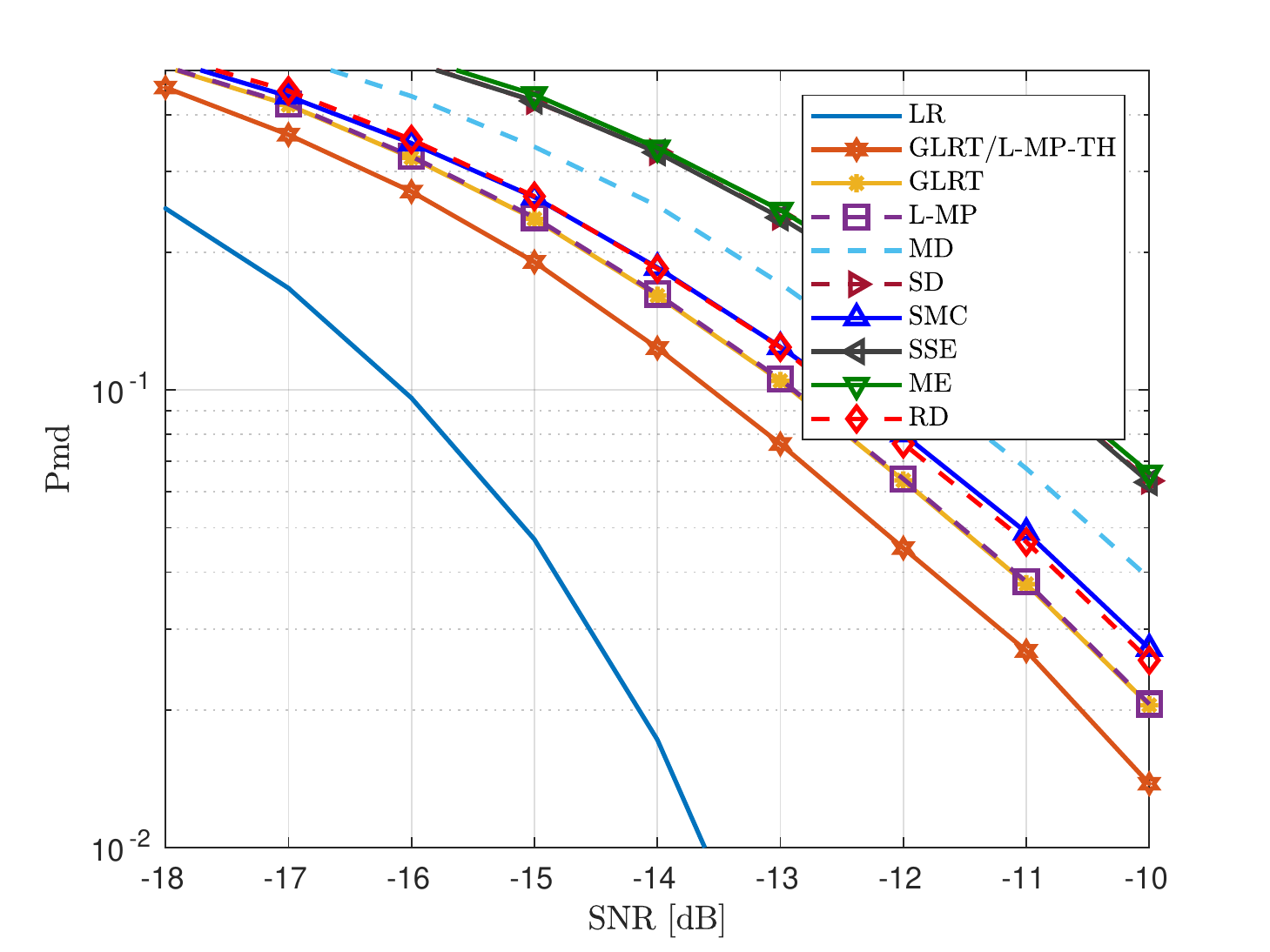}}
	\caption{Miss-detection probability vs SNR when the source is circularly-symmetric complex Gaussian distributed with the indicated $M$ and $L$, and $P_\text{fa}\leq 0.1$.}
	\label{fig:PdVsSNR}
\end{figure}

In Figs. \ref{fig:crocs} and \ref{fig:PdVsSNR}, we show the complementary receiver operating characteristic (CROC) of the detectors, and the miss-detection probability of the detectors against the SNR (with $P_\text{fa}\leq 0.1$), respectively, for different combinations of the parameters $M$ and $L$, and when the source is circular complex Gaussian distributed. 
Besides the genie-aided LRT test, the GLRT and the L-MP tests achieve the best performance among all implementable tests in the shown scenarios, and indistinguishably among them (their curves are superimposed). Considering the theoretical results of the previous section, this is expected for the asymptotic scenario, when $L$ grows to infinity, however, we are obtaining that $T_\text{G}$ and $T_\text{L-MP}$ are almost equivalent even for small data size ($L=10$). We also note that the asymptotic theoretical result of both GLRT and L-MP (labeled as GLRT/L-MP-TH) predicts really well the performance of the finite data case. A larger deviation between the curves GLRT/L-MP-TH and GLRT/L-MP (computed through Monte Carlo simulations) occurs when $M$ decreases from $50$ to $20$. This behavior can be explained in terms of the real distribution of the observations $\{\ve z_l\}$ in (\ref{eq:SUenergy}), which, through the CLT theorem, is closer to the multivariate Gaussian distribution for higher values of $M$, and converges to it (in distribution) when $M\rightarrow\infty$. 

\begin{figure}
\subfigure[Complementary receiver operating characteristics when $\text{SNR}=-11$ dB.] {\includegraphics[width=.85\linewidth]{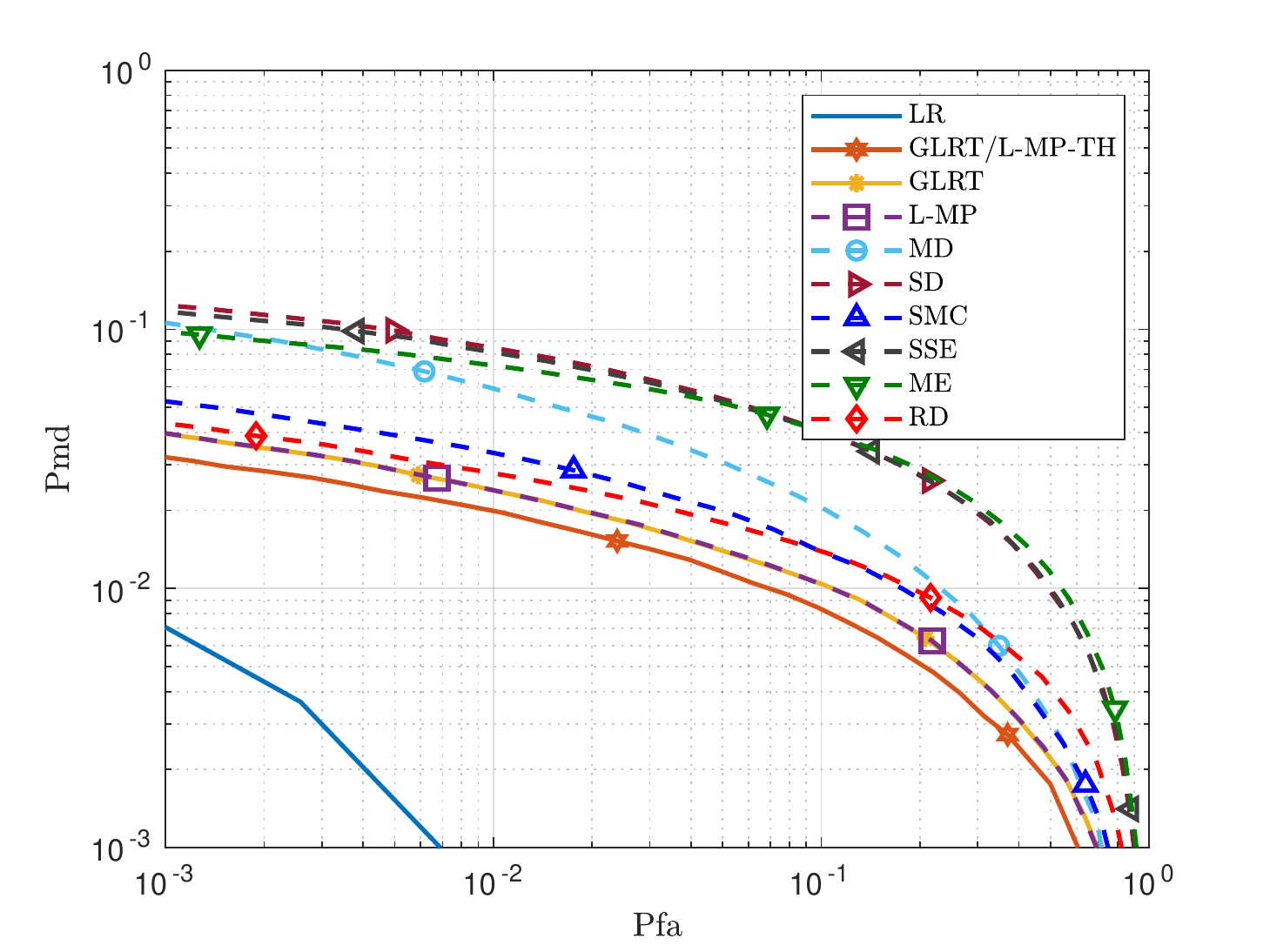}}
\subfigure[Miss-detection probabilities vs SNR when $P_\text{fa}=0.1$.] {\includegraphics[width=.85\linewidth]{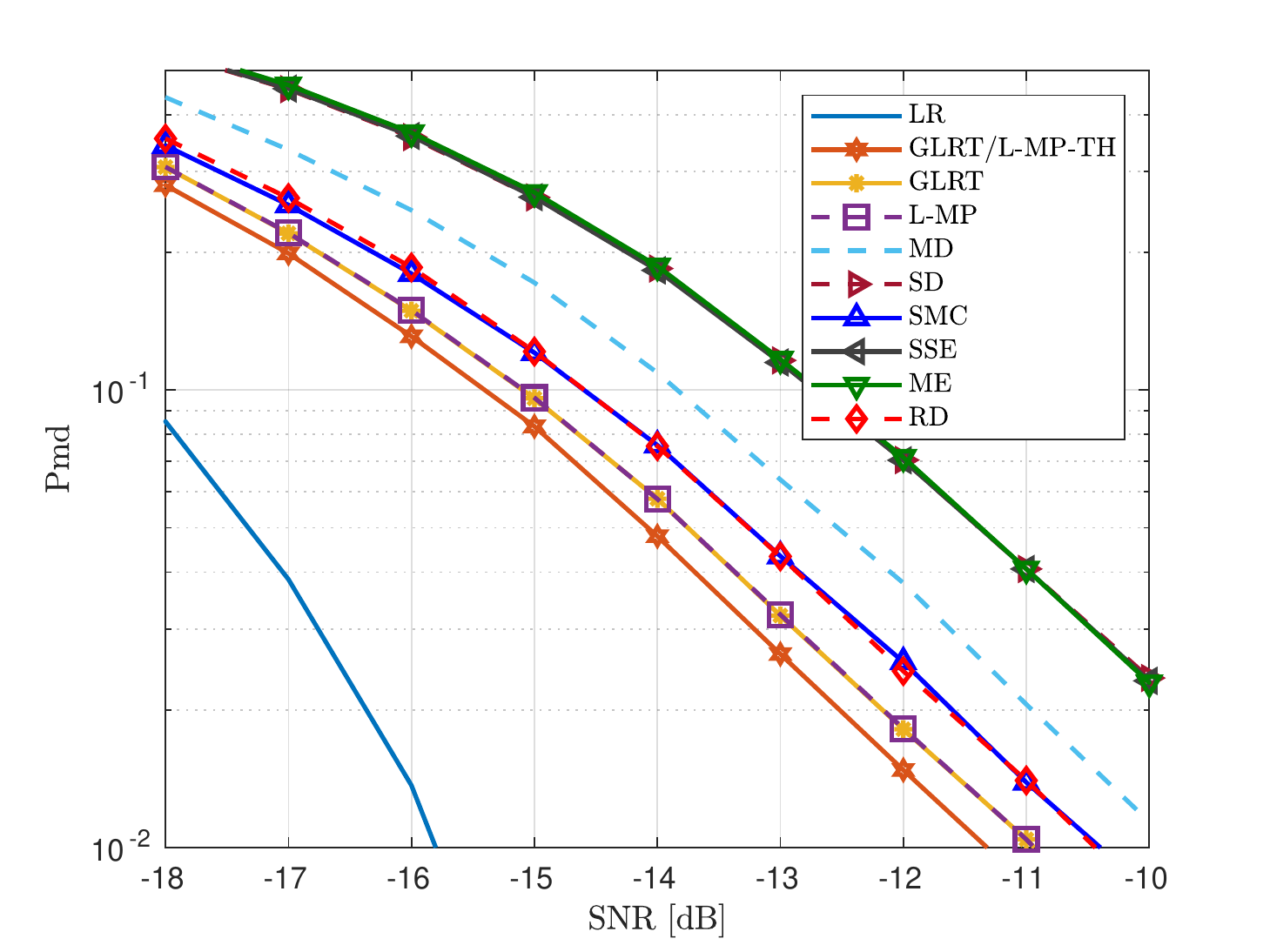}}
\caption{Results when the source is 16-QAM distributed with $M=50$ and $L=10$.}
	\label{fig:qam}
\end{figure}

Observe that the Rao test is known to be asymptotically equivalent to the GLRT, when the MLE is unrestricted, which is not the case considered here. Its performance degradation with respect to the GLRT and L-MP can be at least partially attributed to the fact of not considering the positive constraint on the parameters in the derivation of the statistic. The eigenvalues-based detectors (SSE and ME) could potentially benefit from the correlation of the observations of different nodes (i.e., the term $\ve\theta\ve\theta^T$ in (\ref{eq:meanCoveq})). However, based on the asymptotically equivalence between the GLRT and the L-MP detectors, and the numerical results shown here, the correlation seems to be a second order effect for the considered detection problem.     
In Fig. \ref{fig:qam}, we show the CROCs and the miss-detection probabilities vs the SNR of the considered detectors, when the distribution of the source is 16-QAM. The results are quite similar to those obtained when the source is circular complex Gaussian distributed (compare with Fig. \ref{fig:crocs}(a) and Fig. \ref{fig:PdVsSNR}(a), also with $M=50$ and $L=10$).

\begin{figure}
	\centering
	\includegraphics[width=\linewidth]{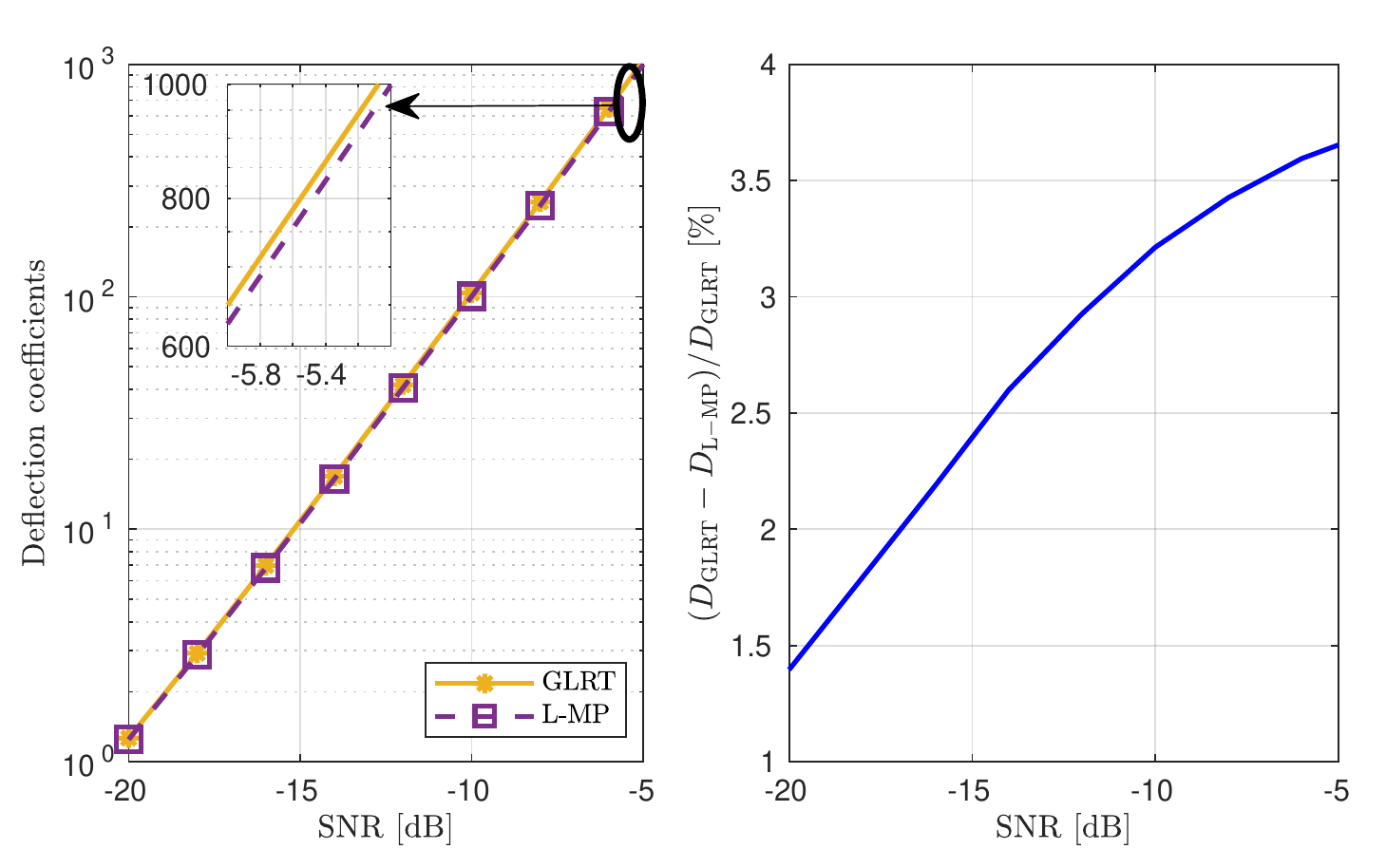}
	\caption{Deflection coefficients of the GLRT and the L-MP (left) and their relative difference (right) for $M=50$ and $L=10$.}
	\label{fig:def}
\end{figure}

Finally, in Fig. \ref{fig:def} (left figure) we show the deflection coefficients for the GLRT and the L-MP computed through Monte Carlo simulations for $M=50$ and $L=10$. On the right side of the same figure, we plot the relative difference between both of them. The deflection coefficient of a statistic is a measure to characterize the detection performance of a test, which is easy to compute and has a monotone correspondence with the ultimate performance (the higher deflection coefficient, the better detection performance) \cite[Sec. 2.4]{Levy_Det}. It is defined as $D_T=|\Ex_1(T)-\Ex_0(T)|/\sqrt{\Var_0(T)}$, where $\Ex_i$ is the expectation operator under $\Hip_i$, $i=0,1$, and $\Var_0$ is the variance operator under $ \Hip_0$. We see that the GLRT performs slightly better than the L-MP in the finite regime ($L=10$), and that the difference between both of them increases with the SNR. Observe also that the SNR can be expressed as follows in terms of the vector parameter $\ve\theta_1$: $\text{SNR}=\Ex(\tfrac{1}{N}\sum_{i=1}^N \theta_n)$, where the expectation is over the fading channels, $h_n$. Therefore, the higher the SNR, the higher impact of the observations' correlation (see (\ref{eq:meanCoveq})), and thus, the better performance of the GLRT as compared to the L-MP. 
However, the small difference in the deflection coefficients in Fig. \ref{fig:def} has little impact on the error probabilities, as shown in Fig. \ref{fig:crocs} and \ref{fig:PdVsSNR}. GLRT is expected to work substantially better than the L-MP in much higher SNR scenarios, where typically even simple detectors, such as the sample mean detector, perform sufficiently well.   
\section{Concluding remarks}
In this paper, we considered the problem of detecting a stochastic radio source signal in a distributed scenario. When the radio source signal is present, the network measurements are shown to be statistically dependent and follow a PDF parametrized by an unknown $N$-vector parameter,  where $N$ is the amount of nodes of the sensing network. Moreover, the components of this vector are constrained to be positive. Considering the absence of an optimal test, we followed the GLRT approach. Although the restrictions of the model make the characterization of the GLRT more involved than the classical unrestricted GLRT, we derived new theoretical expressions to assess its asymptotic performance, when the amount of measurements at the nodes tend to infinity. 

In addition, we showed that the GLRT is not well suited to distributed scenarios because the spatial correlation present in the network measurements complicates the implementation of the test. Therefore, we studied the L-MP test, where the joint PDF of the observations is replaced by the product of the marginal PDFs, and therefore, the data dependence structure is completely discarded. This statistic is much more efficiently implemented in distributed scenarios than the GLRT, and more importantly, it preserves the GLRT's asymptotic performance. This theoretical result also allows us to conclude that the spatial statistical dependence of the network measurements is irrelevant to the detection problem in the asymptotic scenario. Monte Carlo simulations show similar results in the finite-length data regime.

Future research directions may include evaluating the algorithms with real measurements, considering scenarios with multiple radio sources, and exploring optimal strategies for the joint detection and localization problem of radio sources.

\appendices
\section{Moments of the energy samples}
\label{app:gauss_mom}
We compute the mean and covariance matrix of $\ve z$ under $\Hip_1$. Under $\Hip_0$, they can be obtained by setting  $h_n=0$, $n\in[1,N]$ in the result. 
Equations (\ref{eq:signal_rx}) and (\ref{eq:SUenergy}) can be written in vector form as
$\ve y_n = h_n\ve s +\ve v_n$, and ${z}_n = \frac{1}{\sqrt{M}N_0}(\|\ve y_n \|^2-M N_0)$,
where $\ve s = [s(1),\dots,s(M)]^T$, and $\ve v_n = [v_n(1),\dots,v_n(M)]^T$. Then,
\begin{align}
\Ex({z}_n) \!&=\! \frac{1}{\sqrt{M}N_0}\Ex(\|h_n\ve s\|^2\! + \|\ve v_n\|^2\! + 2\Re(h_n^*\ve s^H\ve v_n) -M N_0)\nonumber\\
&=\!  \sqrt{M}\frac{E_s |h_n|^2}{N_0},\label{eq:mz}
\end{align}
where we have used that $\Ex(\|\ve s\|^2) = M E_s$, the signal from the source is independent of $\ve v_n$, and $\ve v_n$ is a zero-mean random vector.
On the other hand, the covariance of ${z}_n$ and ${z}_{n'}$, with $n,n'\in[1,N]$ is
\begin{align}
\text{Cov}({z}_n,{z}_{n'})= \frac{1}{M N_0^2}\Ex\big[(\alpha_n + \beta_n + \gamma_n) (\alpha_{n'} + \beta_{n'} + \gamma_{n'})\big] \label{eq:covz}
\end{align} 
where $\alpha_n = \|h_n\ve s\|^2 - |h_n|^2 M E_s $, $\beta_n = \|\ve v_n\|^2 - M N_0$ and $\gamma_n=2\Re( h_n^* \ve s^H\ve v_n)$ are all zero-mean random variables. Then,
\begin{align*}
\Ex(\alpha_n\alpha_{n'}) &= M |h_{n}|^2 |h_{n'}|^2 E_{s}^2,\\
\Ex(\beta_n\beta_{n'})&= M N_0^2 \delta_{nn'},\\
\Ex(\gamma_n\gamma_{n'})&= 2 M N_0 E_s |h_n|^2\delta_{nn'},
\end{align*}
where $\delta_{nn'}$ is the Kronecker delta. 
The rest of the terms are zero: $\Ex(\alpha_n\beta_{n'})=0$ because $\alpha_n$ and $\beta_{n'}$ are zero-mean and independent random variables;  $\Ex(\alpha_n\gamma_{n'})=0$ because  $\ve s$ and $\ve v_{n'}$ are independent and $\ve v_{n'}$ is a zero-mean random vector; and $\Ex(\beta_n\gamma_{n'})=0$ because $\ve s$ and $\ve v_{n'}$ are independent and $\ve s$ is a zero-mean random vector. Finally, we use (\ref{eq:mz}) and (\ref{eq:covz}) to obtain (\ref{eq:htpeq}).

\section{Restricted GLRT's proofs}
\subsection{Proof of Properties \ref{lemma:1}}
\label{app:prop_pdf}	
P1) We start with the first property. Let $\ve z\in\R^N$. From \cite[eq. (3C.5)]{Kay_SSP_ET}, we have that
\begin{align*}
	\tfrac{\partial\log p(\ve z;\ve\theta)}{\partial\theta_k} = -\tfrac{1}{2}\tr(\mat\Sigma(\ve\theta)\tfrac{\partial\mat\Sigma(\ve\theta)}{\partial\theta_k}) + \tfrac{\partial\ve\mu(\ve\theta)}{\partial\theta_k}\mat\Sigma^{-1}(\ve\theta)(\ve z-\ve\mu(\ve\theta)) \\
	- \tfrac{1}{2} (\ve z-\ve\mu(\ve\theta))^T \tfrac{\partial\mat\Sigma^{-1}(\ve\theta)}{\partial\theta_k} (\ve z-\ve\mu(\ve\theta)).
\end{align*}
We also have that $(\ve\mu(\ve\theta))_i = \sqrt{M}\theta_i$, and that $(\mat\Sigma(\ve\theta))_{i,j} = (1+2\theta_i)\delta_{ij}+\theta_i \theta_j$, for $i,j\in[1,N]$. Then, their derivatives of any order exist for any $\ve\theta$. On the other hand, we need to prove that $\mat\Sigma^{-m}(\ve\theta)$ exists, for (at least) $m=1,2$ and $3$, in a given neighborhood of $\ve\theta = \ve 0$. A sufficient condition for that is $\mat\Sigma(\ve\theta)$ to be positive definite, which is satisfied if $|\theta_i|<\tfrac{1}{\sqrt{N-1}} \ \forall i$.
This condition restricts the definition of the neighborhood $\mathcal{B}$ of $\ve\theta=\ve 0$, and guarantees that $\mat\Sigma(\ve\theta)$ is strictly diagonally dominant. Then, a symmetric diagonally dominant real matrix with nonnegative diagonal entries is positive definite. Therefore, $\mat\Sigma^{-m}(\ve\theta)$ exists for any positive integer $m$. Then, P1) is satisfied.

P2) It is straightforward to see that $\tfrac{\partial^3\log p(\ve z;\ve\theta)}{\partial\theta_k\partial\theta_j \partial\theta_i}$ can be expressed as $\tfrac{\partial^3\log p(\ve z;\ve\theta)}{\partial\theta_k\partial\theta_j \partial\theta_i} = f_1(\ve\theta) + \ve{f}_2(\ve\theta)^T\ve z + \ve z^T \mat{F}_3(\ve\theta)\ve z$, where $f_1(\ve\theta)$, $\ve{f}_2(\ve\theta)$ and $\mat{F}_3(\ve\theta)$ are, respectively, a scalar, vector and matrix bounded functions of $\ve\theta$, if $\ve\theta\in\mathcal{B}$. In addition, the expectation under $p(\ve z; \ve\theta)$ of the linear and quadratic terms of $\ve z$ are also bounded, if $\ve\theta\in\mathcal{B}$. Therefore, the second property follows.

P3) is clearly satisfied given that $\ve i(\ve 0) = (M+2)\mat I_N$, as shown in App. \ref{app:FIM}.

\subsection{Proof of Lemma \ref{lemma:2}}
\label{app:ad_glrt}
Using Theorem 1 from \cite{chernoff1954distribution}, the asymptotic distribution of $2\log T_\text{G}(\ve z_{1:L})$ under $\Hip_0$ is precisely the distribution of 
\begin{equation}
	\ve u^T \ve i(\ve 0_N)\ve u - \inf_{\ve\theta:\ve\theta\succeq \ve 0_N} (\ve u -\ve\theta)^T \ve i(\ve 0_N)(\ve u-\ve\theta),
	\label{eq:dist_H0}
\end{equation}
where $\ve u\sim \N(\ve 0_N,\ve i(\ve 0_N)^{-1})$. As can be seen in the next section, $\ve i(\ve 0_N)= (M+2)\mat I_N$. Replacing this in (\ref{eq:dist_H0}), using that $\tilde{\ve u}\defeq \ve i (\ve 0_N)^{\frac{1}{2}} \ve u \sim\N(\ve 0_N,\mat I_N)$, and observing that a positive scaling factor does not affect the optimization set, we have
\begin{align}
	2\log T_\text{G}(\ve z_{1:L})\overset{a}{\sim} \|\tilde{\ve u}\|^2 -\!\! \inf_{\ve\theta:\ve\theta\succeq \ve 0_N} \|\tilde{\ve u} -\ve\theta\|^2 =\!\sum_{i=1}^N  \tilde{u}_i^2 \bigone(\tilde{u}_i\!\geq 0), \label{eq:dist_H0_b}
\end{align}
where $\tilde{u}_i$ is the $i$-th component of $\tilde{\ve u}$. We see that the asymptotic distribution of the GLRT under $\Hip_0$ is (\ref{eq:chi2}), i.e., the distribution of the sum of $N$ independent truncated central chi-square random variables with one degree of freedom. 
Then, it is straightforward to show that the CDF and the characteristic function (CF) of $\tilde{u}_i^2 \bigone(\tilde{u}_i\geq 0)$ are, respectively, $\Phi(\sqrt{t})\bigone(t\geq 0)$ and $\frac{1}{2}\big(1+\frac{1}{(1-2\imath w)^\frac{1}{2}}\big)$, where $\Phi$ is the standard normal CDF. Then, as $\tilde{u}_i$, $i\in[1,N]$ are i.i.d., the CF of $\sum_{i=1}^N\tilde{u}_i^2 \bigone(\tilde{u}_i\geq 0)$ is $\frac{1}{2^N}\big(1+\frac{1}{(1-2\imath w)^\frac{1}{2}}\big)^N$. Finally, by applying the binomial theorem, and the inverse Fourier transform, we obtain the CDF in (\ref{eq:CDF_H0}). 

Now we characterize the asymptotic distribution of the restricted GLRT under $\Hip_1$. Using P1) of Properties 1, consider the following Taylor expansion
\begin{align}
	\log p(\ve z_{1:L};\ve\theta) &= \log p(\ve z_{1:L};\ve 0) + \ve a(\ve z_{1:L})^T \ve \theta - \tfrac{1}{2}\ve\theta^T \mat B(\ve z_{1:L})\ve\theta \nonumber\\
	& + \|\ve\theta\|^3\mathcal{O}_p(L),\label{eq:maxim}
\end{align}
with the following definitions: $\ve a(\ve z_{1:L})\defeq \tfrac{\partial\log p(\ve z_{1:L};\ve 0)}{\partial\ve\theta}\in\R^N$, and $\mat B(\ve z_{1:L})\defeq -\tfrac{\partial^2\log p(\ve z_{1:L};\ve 0)}{\partial\ve\theta \partial\ve\theta^T}\in\R^{N\times N}$.
The last term of the previous equation is due to the fact that each of the third-order derivatives of $p(\ve z;\ve\theta)$ w.r.t. $\theta_i$, $\theta_j$ and $\theta_k$ $\forall i,j,k$ converges in probability to a constant if they are normalized by the factor $1/L$. Here we used the property P2) of Properties 1.
Then, (\ref{eq:glrt}) can be equivalently written as
\begin{align}
	\log T_\text{G}(\ve z_{1:L}) &= \max_{\ve \theta\succeq\ve 0_N} \log p(\ve z_{1:L};\ve\theta) - \log p(\ve z_{1:L};\ve 0)\nonumber\\
	&= \max_{\ve \theta\succeq\ve 0_N} \ve a(\ve z_{1:L})^T \ve \theta - \tfrac{1}{2}\ve\theta^T \mat B(\ve z_{1:L})\ve\theta \nonumber \\ 
	&+ \|\ve\theta\|^3\mathcal{O}_p(L).\label{eq:glrt2}
\end{align}
Consider that the hypothesis $\Hip_1$ is true, and consider also the weak signal assumption. Then, the first term of (\ref{eq:glrt2}) converges in distribution to a normal random variable (c.f. (\ref{eq:inner_dist})). Second, we have that \cite[App. 7B]{Kay_SSP_ET} $\tfrac{1}{L}\mat B(\ve z_{1:L})\rightarrow \ve i(\ve 0) = (M+2) \mat I_N$ in probability, as $L\rightarrow\infty$. Thus, $\ve\theta^T \mat B(\ve z_{1:L})\ve\theta = \mathcal{O}_p(1)$. Additionally, the third term of (\ref{eq:glrt2}) converges to zero as $\mathcal{O}_p(L^{-1/2})$. Therefore, when $L$ is sufficiently large, we can solve the above optimization problem considering only the first two terms of (\ref{eq:glrt2}). The fact that $\ve i(\ve 0)$ is a diagonal matrix makes the computation of the restricted GLRT easier given that

\begin{align}
	\log T_\text{G}(\ve z_{1:L}) &= \sum_{n=1}^N \max_{\theta_i\geq 0} -\tfrac{1}{2} B_{ii}(\ve z_{1:L}) \big(\theta_i - \tfrac{a_i(\ve z_{1:L})}{B_{ii}(\ve z_{1:L})}\big)^2 \nonumber\\
	&+ \tfrac{1}{2} B_{ii}(\ve z_{1:L})\big(\tfrac{a_i(\ve z_{1:L})}{B_{ii}(\ve z_{1:L})}\big)^2 + \mathcal{O}_p(L^{-1/2}) \label{eq:glrt3}
\end{align}
In the last equation, $a_i(\ve z_{1:L})$ and $B_{ii}(\ve z_{1:L})$ are the $i$-th and $(i,i)$-th element of $\ve a(\ve z_{1:L})$ and $\mat B(\ve z_{1:L})$, respectively. For $L$ large enough, $\tfrac{1}{L}B_{ii}(\ve z_{1:L})>0$ (c.f. P3) in Properties 1). Therefore, the $i$-th element of the positive constrained MLE is $\hat{\theta}_i = \big\{\frac{a_i(\ve z_{1:L})}{B_{ii}(\ve z_{1:L})}\big\}_+$, $i\in[1,N]$. Then, $(\ref{eq:glrt3})$ becomes 
\begin{align}
	\log T_\text{G}(\ve z_{1:L}) &= \frac{1}{2}\sum_{n=1}^N \left\{\frac{a_i(\ve z_{1:L})}{\sqrt{B_{ii}(\ve z_{1:L})}}\right\}_+^2 + \mathcal{O}_p(L^{-1/2}) \nonumber\\
	&= \frac{1}{2} \left\| \left\{(\tfrac{1}{L}\mat B(\ve z_{1:L}))^{-1/2} \tfrac{1}{\sqrt{L}} \ve a(\ve z_{1:L}) \right\}_+\right\|^2 \nonumber\\
	&+ \mathcal{O}_p(L^{-1/2}). \label{eq:glrt4}
\end{align}
Using  P3) in Properties 1 and the classical results of the GLRT theory \cite[App. 7B]{Kay_SSP_ET} (i.e., the convergence in distribution of $\tfrac{1}{\sqrt{L}} \ve a(\ve z_{1:L})$ via the central limit theorem and the continuous mapping theorem), we have that
\begin{align}
	(\tfrac{1}{L}\mat B(\ve z_{1:L}))^{-1/2} \tfrac{1}{\sqrt{L}} \ve a(\ve z_{1:L})&\overset{a}{\sim}\N(\sqrt{L} \ve i(\ve 0)^{1/2}\ve\theta_1,\mat I_N)\nonumber\\
	&\overset{a}{\sim}\N(\sqrt{L(M+2)} \ve\theta_1,\mat I_N) \label{eq:inner_dist}
\end{align}
Using (\ref{eq:inner_dist}), it is clear that the asymptotic distribution of (\ref{eq:glrt4}) under $\Hip_1$ is (\ref{eq:chi2}), i.e., the sum of $N$ independent non-central truncated chi-square random variables with one degree of freedom. Unfortunately, this distribution has not a closed-form expression. However, it is actually possible to get its characteristic function. From \cite[Lemma 2, eq. (23)]{maya2023selection}, we obtain (\ref{eq:chi2trunc}).

\subsection{Computation of the Fisher information matrix}
\label{app:FIM}
When $p(\ve z;\ve \theta)$ is Gaussian, as in (\ref{eq:htpeq}), the $i,j$-th component of the Fisher information matrix can be computed as \cite{Kay_SSP_ET}
\begin{align}
	[\ve i(\ve\theta)]_{i,j} &=\frac{\partial\ve\mu(\ve\theta)}{\partial\theta_i}\mathbf\Sigma(\ve\theta)^{-1}\frac{\partial\ve\mu(\ve\theta)}{\partial\theta_j}\nonumber\\
	&+\frac{1}{2}\tr\left[\mathbf\Sigma(\ve\theta)^{-1}\frac{\partial\mathbf\Sigma(\ve\theta)}{\partial\theta_i} \mathbf\Sigma(\ve\theta)^{-1} \frac{\partial\mathbf\Sigma(\ve\theta)}{\partial\theta_j}\right]. \label{eq:FIM_Guass}
\end{align}
From (\ref{eq:meanCoveq}), we have $\frac{\partial\ve{\mu}(\ve \theta)}{\partial\theta_i} = \sqrt{M}\ve e_i$, and $\frac{\partial\mathbf\Sigma(\ve \theta)}{\partial\theta_i} = (\ve\theta \ve e_i^T + \ve e_i \ve\theta^T)+ 2\ve e_i\ve e_i^T$, where $\ve e_i\in\R^N$ is the canonical vector with value $1$ in the $i$-th coordinate and $0$ otherwise. Then, $\ve i(\ve 0_N)= (M+2)\mat I_N$.

\section{Restricted L-MP's proofs}
\subsection{Proof of Lemma \ref{lemma:L-MP}}
\label{app:L-MP}
In order to characterize the asymptotic distribution of $T_\text{L-MP}$ under $\Hip_0$, we can apply again Theorem 1 in \cite{chernoff1954distribution} separately to each of the $N$ scalar optimization problems in (\ref{eq:LMP}). Then, the asymptotic distribution of $2\log T_\text{L-MP} (\ve z_{1:L})$ under $\Hip_0$ coincides with the distribution of  
\begin{equation}
	\sum_{k=1}^{N} \left(j_k(0)u_k^2 - \inf_{\theta_k:\theta_k\geq 0} j_k(0)(u_k -\theta_k)^2\right),\label{eq:dist_LMP_H0}
\end{equation}
where $u_k\sim\N(0,j_k(0)^{-1})$ with $k\in[1,N]$ are independent random variables, and $j_k(0)=\Ex\big(\big(\frac{\partial \log p_k(z_k;0)}{\partial \theta_k}\big)^2\big)$. Using (\ref{eq:FIM_Guass}), we have $j_k(0)=M+2$ for all $k$. Then, it is evident that the distribution of (\ref{eq:dist_LMP_H0}) and (\ref{eq:dist_H0_b}) is the same, proving that the asymptotic distribution of $T_\text{G}$ and $T_\text{L-MP}$ under $\Hip_0$ are equal.

Now it remains to find the distribution of the restricted $T_\text{L-MP}$ under $\Hip_1$. Consider the following definitions: $p_\text{MP}(\ve z_{1:L};\ve\theta)\defeq\prod_{k=1}^N\prod_{l=1}^L p_k(z_k(l);\theta_k)$, $\tilde{\ve a}(\ve z_{1:L})\defeq \tfrac{\partial\log p_\text{MP}(\ve z_{1:L};\ve 0)}{\partial\ve\theta}\in\R^N$, and $\tilde{\mat B}(\ve z_{1:L})\defeq -\tfrac{\partial^2\log p_\text{MP}(\ve z_{1:L};\ve 0)}{\partial\ve\theta \partial\ve\theta^T}\in\R^{N\times N}$. It is straightforward to show that also in this case $\tfrac{1}{L}\tilde{\mat B}(\ve z_{1:L})\rightarrow \ve j(\ve 0) = (M+2)\mat I_N$, in probability, where $\ve j(\ve 0)$ is defined in (\ref{eq:j}). Using the same arguments than in the previous section, we can express the restricted L-MP as follows: 
\begin{align}
	\log T_\text{L-MP}(\ve z_{1:L}) &= \max_{\ve \theta\succeq\ve 0_N} \log p_\text{MP}(\ve z_{1:L};\ve\theta) - \log p_\text{MP}(\ve z_{1:L};\ve 0)\nonumber\\
	&= \max_{\ve \theta\succeq\ve 0_N} \tilde{\ve a}(\ve z_{1:L})^T \ve \theta - \tfrac{1}{2}\ve\theta^T \tilde{\mat B}(\ve z_{1:L})\ve\theta \nonumber \\ 
	&+ \|\ve\theta\|^3\mathcal{O}_p(L).\nonumber\\
	&= \frac{1}{2} \left\| \left\{(\tfrac{1}{L}\tilde{\mat B}(\ve z_{1:L}))^{-1/2} \tfrac{1}{\sqrt{L}} \tilde{\ve a}(\ve z_{1:L}) \right\}_+\right\|^2 \nonumber\\
	&+ \mathcal{O}_p(L^{-1/2}). \label{eq:clmp}
\end{align}
From the asymptotic distribution of the \emph{unrestricted} L-MP under $\Hip_1$ in (\ref{eq:LMP1}) (see next section), we have that
\begin{align}
	(\tfrac{1}{L}\tilde{\mat B}(\ve z_{1:L}))^{-1/2} \tfrac{1}{\sqrt{L}} \tilde{\ve a}(\ve z_{1:L}) \overset{a}{\sim}\N(\sqrt{L(M+2)}\ve \theta_1,\mat I_N). \label{eq:Batilde}
\end{align}	
Now, comparing (\ref{eq:clmp}) and (\ref{eq:Batilde}) with (\ref{eq:glrt4}) and (\ref{eq:inner_dist}), we conclude that the asymptotic distribution of the restricted L-MP is exactly the same as the restricted GLRT.

\subsection{Asymptotic distribution of the unrestricted L-MP under $\Hip_1$}
\label{asymp_dist}
The following results are used to prove the asymptotic distribution of the unrestricted $T_\text{L-MP}$ (actually $2\log T_\text{L-MP}$) under $\Hip_1$. The proof is found in \cite{maya2021distributed}. 
\begin{lemma}
	\label{lem:localparam}
	Assume 
	i) the first and second-order derivatives of the log-likelihood function are well defined and continuous functions.
	ii) $\Ex[\partial\log p_k(z_k(l);\theta_k)/\partial\theta_k]\!= 0$, $\forall l$, $k\!\in[1,N]$.
	iii) the matrix $\ve j( \ve \theta_1)$ defined in (\ref{eq:j}) is nonsingular.
	Then, the asymptotic distribution of $T_\text{L-MP}$ under $\Hip_1$ is:
	\begin{align}
		2\log T_{\text{L-MP}}(\ve z) & \overset{a}{\sim} g_N(\ve \mu_{ \text{MP},1},\ve{\Sigma}_{\text{MP},1}),
		\label{eq:mpglr}
	\end{align}
	where $\ve\mu_{\text{MP},1} = \sqrt{L} \ve i_\text{MP}(\ve\theta_1)^{\frac{1}{2}} (\ve\theta_1\!\!-\ve\theta_0)$,~and
	\noindent $\ve\Sigma_{\text{MP},i} \!\!=\! \ve i_\text{MP}(\ve\theta_i)^{\frac{1}{2}} \ve j(\ve \theta_i)^{-1} \tilde{\ve i}(\ve \theta_i) \ve j(\ve \theta_i)^{-1}  \ve i_\text{MP}(\ve\theta_i)^{\frac{1}{2}} $, $i=0,1$, $p_\text{MP}(\ve z_l; \ve \theta_1) \defeq\Pi_{k=1}^N p_k(z_k(l); \theta_{1k})$ is the product of the marginal PDFs used for building $T_\text{L-MP}$, and
	\begin{align} 
		&\textstyle [\tilde{\ve i}(\ve \theta_1)]_{kj} \hspace*{-2.5pt}\defeq\Ex_{\ve\theta_1}\hspace*{-2pt}\left(\frac{\partial\log p_k(z_k(l);\theta_{1k})} {\partial\theta_k} \frac{\partial\log p_j(z_j(l);\theta_{1j})}{\partial\theta_j}\hspace*{-2pt}\right),
		\label{eq:FisherLocal}\\
		&\textstyle [\ve j(\ve \theta_1)]_{kj} \defeq -\Ex_{\ve\theta_1}\left( \frac{\partial^2 \log p_k(z_k(l);\theta_{1k})} {\partial\theta_k\partial\theta_{j}} \right),\label{eq:j}\\
		& \textstyle \ve i_\text{MP}(\ve\theta_1)\defeq \Ex_{\ve\theta_1}\Big(\frac{\partial\log p_\text{MP}(\ve z_l;\ve\theta_1)}{\partial\ve{\theta}} \frac{\partial\log p_\text{MP}(\ve z_l;\ve{\theta_1})^T}{\partial\ve{\theta}} \Big),
	\end{align}
	where the expectations are taken with respect to $p(\ve z;\ve\theta_1)$.
	We also define $g_N(\ve\mu,\ve\Sigma)$ as the PDF of $\|\ve n\|^2$ when $\ve n\sim\N(\ve\mu,\ve\Sigma)$.
\end{lemma}
\begin{corollary}
	If the signal to be tested is weak, i.e., there exists a constant $c$ such that $\|\ve\theta_1-\ve\theta_0\|\leq\frac{c}{\sqrt{L}}$, then, as $L\rightarrow\infty$,
	\begin{equation}
		\|\ve \mu_{\text{MP},1}\|^2 = L (\ve\theta_1-\ve\theta_0)^T \ve i_\text{MP}(\ve\theta_0)(\ve\theta_1-\ve\theta_0),
		\label{eq:mu_mp1}
	\end{equation}
	and $\ve\Sigma_{\text{MP},1} = \ve\Sigma_{\text{MP},0}$.
\end{corollary}

Now we evaluate the quantities defined in the previous results for $p(\ve z;\ve\theta)$ in (\ref{eq:htpeq}) under the weak signal hypothesis.
It is straightforward to show that $\tilde{\ve i}(\ve 0_N) = \ve i_\text{MP}(\ve 0_N) = \ve j(\ve 0_N)=(M+2)\mat I_N$. Thus, $\ve\Sigma_{\text{MP},1} = \ve\Sigma_{\text{MP},0}=\mat I_N$. Using (\ref{eq:mu_mp1}), we obtain $\|\ve \mu_{\text{MP},1}\|^2 =L(M+2)\|\ve \theta_1\|^2$. Finally, the asymptotic distribution of the unrestricted L-MP is 
\begin{align}
	2\log T_{\text{L-MP}}(\ve z) \overset{a}{\sim} g_N(L(M+2)\|\ve \theta_1\|^2, \mat I_N)\sim \|\ve n\|^2,\label{eq:LMP1}
\end{align}
where $\ve n\sim\N(\sqrt{L(M+2)}\ve \theta_1,\mat I_N)$.

\subsection{Derivation of the local MLE}
\label{app:MLEloc}
Consider sensor $k$ and its energy measurements denoted by  $\ve z^k = [z_k(1),\dots,z_k(L)]^T$. The distribution of $\ve z^k$ under $\Hip_1$ is $\ve z^k \sim \N(\sqrt{M} \theta_k\ve 1_L,(\theta_k+1)^2 \mathbf{I}_L)$. The log-likelihood $\log L_k(\ve z^k;\theta_k)$ is given by (neglecting terms which do not depend on $\theta_k$):
\[\log p(\ve z^k; \theta_k) \propto -L\log(1+\theta_k)-\frac{1}{2}\left\|\frac{\ve z^k-\sqrt{M}\theta_k \ve 1_L}{1+\theta_k}\right\|^2.\]
Deriving this expression with respect to $\theta_k$ and equaling it to zero, it is easy to show that the solution for the MLE has to satisfy the following quadratic equation as long as at least one root is positive (i.e. it belongs to the optimization set): 
\begin{multline}
\theta_k^2+(M+2+\sqrt{M} m^1_k )\theta_k 
-(m^2_k
+\sqrt{M}m^1_k -1) = 0.\nonumber
\end{multline}
If both roots are negative, it can be shown that the log-likelihood is decreasing as a function of $\theta_k$ in the interval $[0,\infty)$, meaning that the MLE is $0$ for that case.
If one root is positive and the other is negative, the positive root is the solution (i.e., the one with positive radical).
It also can be shown that it is not possible to have both roots positive. Thus, the local MLE can be expressed as (\ref{eq:mlelocal}).


\end{document}